\documentclass[twocolumn,superscriptaddress,showpacs,amsmath,amssymb,pra,nofootinbib,longbibliography]{revtex4-1}

\usepackage{dcolumn,bm,book tabs,comment,natbib,isomath,slashbox}
\usepackage[pdftex]{graphicx}
\usepackage[usenames,dvipsnames]{color}
\usepackage[colorinlistoftodos]{todonotes}
\usepackage{colortbl}

\definecolor{URLCOL}{rgb}{0,0.52,0.83} 
\definecolor{LINKCOL}{rgb}{0.05,0.5,0} 
\definecolor{CITECOL}{rgb}{0.25,0,0.48} 

\usepackage[pdftex,bookmarks,breaklinks,bookmarksopen,bookmarksnumbered,colorlinks,linkcolor=LINKCOL,linktocpage,citecolor=CITECOL,urlcolor=URLCOL,pdfpagemode=UseOutline,pdftex]{hyperref}

\definecolor{TITLECOL}{rgb}{0.1,0.2,0.7} 
\definecolor{SECOL}{rgb}{0.1,0.2,0.7} 
\definecolor{CONTENTSCOL}{rgb}{0.1,0.2,0.7} 
\definecolor{SSECOL}{rgb}{0.25,0,0.48} 
\definecolor{SSSECOL}{rgb}{0.2,0.08,0.53} 
\definecolor{FINCOL}{rgb}{0.01,0.3,0.07}

\def\coltableofcontents{ 
	{
		\definecolor{SECOL}{rgb}{0.25,0,0.48} 
		\definecolor{SSECOL}{rgb}{0.2,0.08,0.53} 
		\tableofcontents
	}
}

\def\coloredtitle#1{\title{\textcolor{TITLECOL}{#1}}} 
\def\coloredauthor#1{\author{\textcolor{CITECOL}{#1}}} 

\definecolor{URLCOL}{rgb}{0,0.17,0.43} 
\definecolor{LINKCOL}{rgb}{0.05,0.4,0} 
\definecolor{CITECOL}{rgb}{0.35,0,0.48} 

\def\Tabref#1{Table~\ref{#1}}

\def\Eqref#1{Eq.~\eqref{#1}}

\def\Figref#1{Fig.~\ref{#1}}
\def\Ref#1{Ref.~\cite{#1}}

\def\sec#1{\section{\textcolor{SECOL}{#1}}}
\def\ssec#1{\subsection{\textcolor{SSECOL}{#1}}}

\definecolor{lightgray}{gray}{0.8}

\def\bea{\begin{eqnarray}}
\def\eea{\end{eqnarray}}
\def\ben{\begin{equation}}
\def\een{\end{equation}}
\def\benu{\begin{enumerate}}
\def\enu{\end{enumerate}}

\def\bei{\begin{itemize}}
\def\eei{\end{itemize}}
\def\beit{\begin{itemize}}
\def\eit{\end{itemize}}
\def\benu{\begin{enumerate}}
\def\enu{\end{enumerate}}

\def\sss{\scriptscriptstyle\rm}

\def\1var{(\bx_1...\bx\N)}

\def\br{{\bf r}}

\def\bx{{x}}

\def\ba{{\bf a}}

\def\s{_{\sss S}}

\def\N{_{\sss N}}

\def\loc{^{\rm loc}}

\def\sph_int{ {\int d^3 r}}

\def\figwidth{7cm}

\def\bn{\vectorsym{n}}

\def\bx{\vectorsym{x}}
\def\Tv{\vectorsym{T}}
\def\ba{\vectorsym{\alpha}}

\def\K{\matrixsym{K}}

\def\W{{\rm ^W}}

\def\T{^\top}

\def\K{\matrixsym{K}}

\def\u{\vectorsym{u}}

\def\u{\vectorsym{u}}

\def\loc{^{\rm loc}}
\def\M{\mathcal{M}}
\def\J{\mathcal{J}}
\def\ML{^{\rm ML}}

\def\PCA{^{\scriptsize\rm PCA}}
\def\T{\mathcal{T}}

\begin{document}

\coloredtitle{
Understanding Machine-learned Density Functionals
}
\coloredauthor{Li Li}
\affiliation{Department of Physics and Astronomy, 
University of California, Irvine, CA 92697}

\coloredauthor{John C.\ Snyder}
\affiliation{Department of Physics and Astronomy, 
University of California, Irvine, CA 92697}
\affiliation{Department of Chemistry, 
University of California, Irvine, CA 92697}

\coloredauthor{Isabelle M.\ Pelaschier}
\affiliation{Department of Physics,
Vanderbilt University, Nashville, TN 37235, USA}
\affiliation{Department of Physics and Astronomy, 
University of California, Irvine, CA 92697}

\coloredauthor{Jessica Huang}
\affiliation{Department of Chemistry, 
University of California, Irvine, CA 92697}

\coloredauthor{Uma-Naresh Niranjan}
\affiliation{Department of Computer Science, 
University of California, Irvine, CA 92697}

\coloredauthor{Paul Duncan}
\affiliation{Department of Chemistry, 
University of California, Irvine, CA 92697}

\coloredauthor{Matthias Rupp}
\affiliation{Department of Chemistry, University of Basel, Klingelbergstr.~80, 4056~Basel, Switzerland}

\coloredauthor{Klaus-Robert M{\"u}ller}
\affiliation{Machine Learning Group, Technical University of Berlin, 10587 Berlin, Germany}
\affiliation{Department of Brain and Cognitive Engineering, Korea University,
Anam-dong, Seongbuk-gu, Seoul 136-713, Korea}

\coloredauthor{Kieron Burke}
\affiliation{Department of Chemistry, 
University of California, Irvine, CA 92697}
\affiliation{Department of Physics and Astronomy, 
University of California, Irvine, CA 92697}

\date{\today}

\begin{abstract}
Kernel ridge regression is used to approximate the kinetic energy of non-interacting fermions
in a one-dimensional box as a functional of their density. The properties of different
kernels and methods of cross-validation are explored, and highly accurate
energies are achieved. Accurate {\em constrained optimal densities}
are found via a modified Euler-Lagrange constrained minimization of the total energy.
A projected gradient descent algorithm is derived using local principal component analysis.
Additionally, a sparse grid representation of the density can be used without degrading the
performance of the methods. The implications for machine-learned density
functional approximations are discussed.
\end{abstract}


\maketitle
\coltableofcontents

\sec{Introduction}

Since the early days of quantum mechanics, it has been known that sufficiently accurate solutions of Schr{\"o}dinger's equation for electrons and nuclei yield good predictions of the properties of solids and molecules~\cite{D29}.
But the Coulomb repulsion between electrons causes the computational cost of solving the Schr\"{o}dinger equation to grow rapidly with the number of electrons, $N$~\cite{K99}.
However, as Hohenberg and Kohn proved in 1964 \cite{HK64}, the one-electron density may be used as the basic variable
of quantum mechanics instead of the wavefunction, greatly reducing the complexity of the computational problem. 
This is called density functional theory (DFT)~\cite{DG90}. 
In principle, the mapping of the Schr\"{o}dinger equation to one with the electron density is exact, but in practice, 
both the
kinetic energy and the energy of the interaction between electrons must be approximated.
In the original Thomas-Fermi theory~\cite{T27,F28}, a local density
functional approximation to the kinetic energy is used. 
However, Thomas-Fermi theory proved unsuitable for chemical and solid-state applications as it does not bind matter~\cite{T62}.
Shortly after the Hohenberg-Kohn theorems, Kohn and Sham (KS) \cite{KS65} found a middle ground 
by mapping the many-body system onto a fictitious system of non-interacting electrons which
reproduce the exact electron density. 
The main reason KS DFT became successful is because the kinetic energy of these non-interacting electrons is an 
excellent approximation to the many-body kinetic energy. Simple approximations to the interaction energy produce
much greater accuracy and reliability compared with the standard orbital-free DFT schemes built on
Thomas-Fermi theory.
However,
the accuracy of the results are still sensitive to the approximation of the exchange-correlation (XC)
functional. 
In the past four decades, there has been extensive research into 
improving density functional XC approximations. 
Development of both empirical and non-empirical functionals require great intuition built on years of experience, as
well as painstaking trial and error~\cite{B88,LYP88,PBE96}.

Despite the great success KS DFT has enjoyed, the computational cost scales as $O(N^3)$, which is much worse than the linear scaling of orbital-free DFT \cite{KT12}. 
Thus, there continues to be strong interest in improving upon existing orbital-free approximations to the kinetic energy~\cite{KJTH09,KT12,TWb02}. 
A sufficiently accurate approximation to $T\s[n]$, the kinetic
energy of KS electrons as a functional of the ground-state density $n(\br)$ would enable
highly accurate orbital-free DFT calculations with the same accuracy as KS DFT at a fraction
of the computational cost. For example, benchmark orbital-free DFT calculations are capable
of treating millions of atoms in metals~\cite{HC09} or proteins in solvent~\cite{HLB08}.
Note that accuracy in $T\s$ beyond that of current XC approximations would be
unnecessary, since all standard orbital-free DFT schemes utilize the KS decomposition of
the energy, so that standard XC approximations developed for KS DFT can be utilized.
However, since $T\s$ is typically comparable to the total energy of the system~\cite{DG90}, an unavoidable problem is that a useful kinetic energy (KE) functional calls for much stricter relative accuracy than XC functionals. 
Additionally, accurate functional derivatives are required because one finds the ground state density by solving
an Euler equation with the approximate kinetic energy functional.
Continued efforts have been made in this research direction, with some notable progress \cite{W35,WC00,KCT13,KJTH09b,KJTH13,TW02,CAT85,GAC96,GAC98,WT92,WGC99,XHSC12}.
For a review of state-of-the-art orbital-free DFT functionals, we refer the reader to \Ref{KT12}.

In DFT, functionals typically fall into two categories. Non-empirical functionals derived from first principles tend to work well across a broad range of systems, and may exhibit systemic errors in treating certain
types of interactions. Semi-empirical functionals introduce parameters that are fitted to standard
data sets, and are typically more accurate with less systematic errors.

Recently, some of us applied machine learning (ML) in a completely new approach
to approximating density functionals \cite{SRHM12,SRHB13}.
In a proof of principle, kernel ridge regression was used to approximate the 
kinetic energy of non-interacting fermions
confined to a 1d box as a functional of the electron density \cite{SRHM12}. 
In that work, a modified
orbital-free DFT scheme was able to produce highly accurate self-consistent densities and energies that were systematically improvable with additional training data.
ML algorithms are capable of learning high-dimensional patterns by non-linear interpolation
between given data. 
These powerful methods have proved to be very successful in many applications~\cite{MMRT01}, including medical diagnoses~\cite{K01}, stock market predictions~\cite{HNW05}, automated text categorization~\cite{S02}, and others.
Recently, ML has been applied to quantum chemistry, including
fast and accurate modeling of molecular atomization energies~\cite{RTML12,HMBF13,MRGV13}, optimizing transition state theory dividing surfaces~\cite{PHSR12}, and calculating bulk crystal properties 
at high temperatures~\cite{BPKC10}.

This new approach to density functional approximation suffers none of the typical challenges
found in traditional approximations, but presents many new ones. First and foremost, ML is
data-driven: reference calculations are needed to build a model for the KE functional.
Since every iteration in a KS DFT calculation provides an electron density and its exact non-interacting
KE, reference data is relatively easy to obtain.
Additionally, the ML approximation (MLA) to the KE may have thousands or millions of parameters
and satisfy none of the standard exact conditions in DFT, such as positivity, scaling, and exactness for a uniform
electron gas. On the other hand, the form of the MLA is completely general and thus directly
approximates the functional itself, suffering none of the
typical issues plagued by standard functionals starting from a local approximation.
For example, some of us recently showed that an MLA for the KE has no problem accurately dissociating
soft-Coulomb diatomics in 1d---a huge challenge for standard approximations \cite{SRHB13}.
However, kernel ridge regression is strictly a method of interpolation. An MLA can only
be used on systems it was designed for.

In this paper, we explore the properties of the MLA derived in \Ref{SRHM12} in greater detail.
In particular, we investigate the use of various kernels and their properties and the efficiency of
various cross validation methods. We discuss the issue of functional derivatives
of the MLA in greater detail, and explain how a modified constraint to the standard Euler
equation enables highly accurate self-consistent densities, or 
{\em constrained optimal densities}, to be found. Additionally, a projected gradient descent
algorithm is derived using local principal component analysis in order to solve the modified Euler equation. 
Finally, we explore the use of a sparse grid representation of the electron density and its
effects on the method.

\sec{Theory and Background}

Throughout this work, we consider only non-interacting same-spin fermions in one-dimension.
Thus, all electron densities $n(x)$ are fully spin-polarized.
Atomic units are used in symbolic equations, but energies are usually presented in $\text{kcal}/\text{mol}$.

\ssec{Model system}

\begin{figure}[tb]
\includegraphics[width=\figwidth]{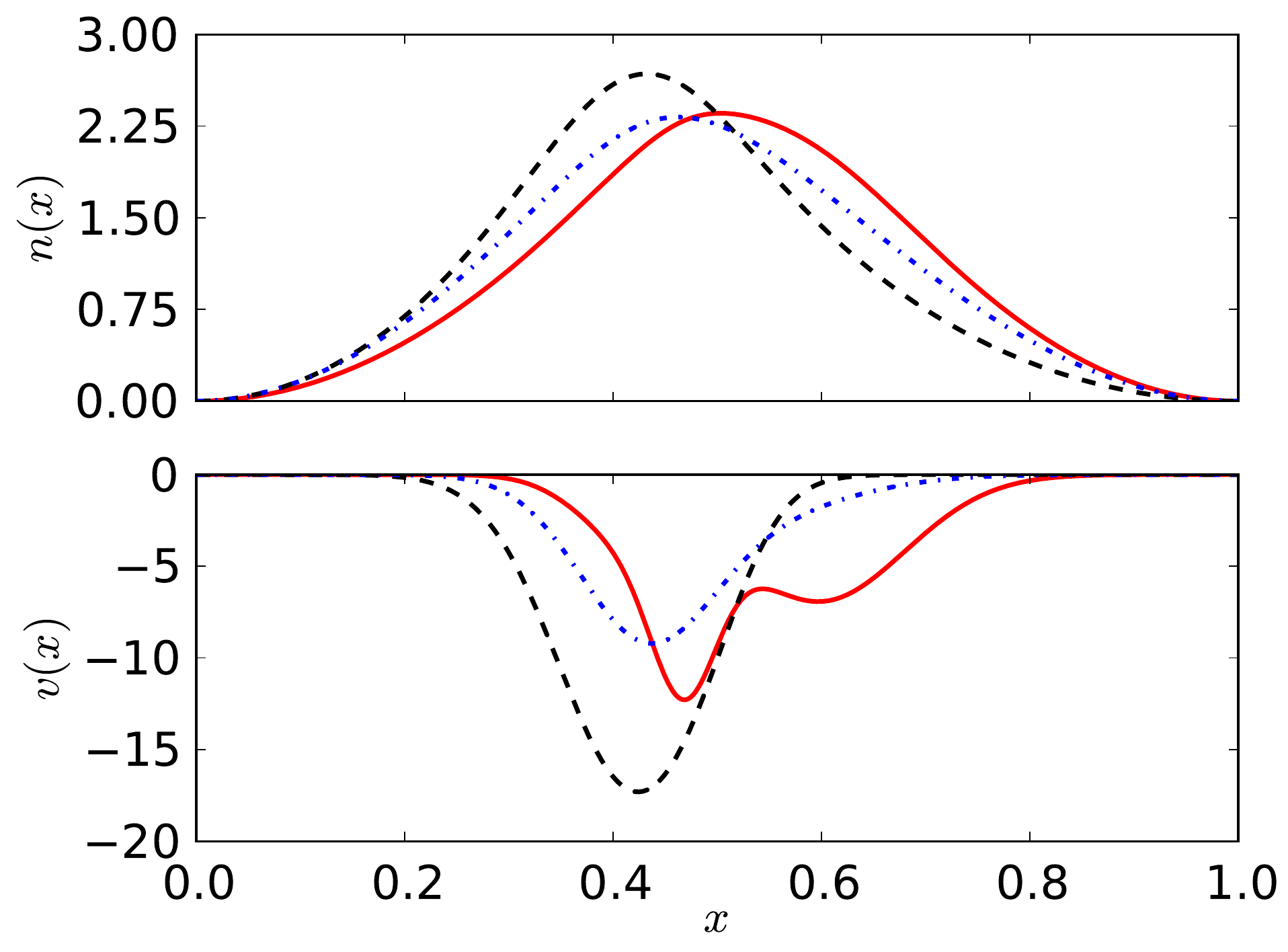}
\caption{A few sample densities and their corresponding potentials, for $N=1$.}
\label{f:data_example}
\end{figure}

Consider $N$ non-interacting same-spin fermions subject to a smooth external potential
in one-dimension, with hard walls at $x=0$ and $x=1$.
We restrict this study to a simple class of potentials, namely a sum of 3 Gaussian dips
with varying heights, widths and centers:
\ben
v(x) = \sum_{j=1}^3 a_j \exp(-(x - b_j)^2/(2 c_j^2)),
\label{eq:potential}
\een
for $x\in[0,1]$, and $v(x)=\infty$ elsewhere.
The Hamiltonian for this system is simply
$\hat H = \hat T + \hat V$,
where 
$\hat T = -\partial^2/2\partial x^2$ and $\hat V = v(x)$. We solve the Schr\"{o}dinger equation
\begin{equation}
\left( -\frac{1}{2} \frac{\partial^2}{\partial x^2} + v (x)\right) \phi (x)=\epsilon \phi (x),
\label{eq:scheq}
\end{equation}
for the eigenvalues $\epsilon_j$ and orbitals $\phi_j(x)$.
As our fermions are same-spin, each orbital $\phi_j(x)$ is singly-occupied. Thus,
the electron density is given by
\ben
n(x) = \sum_{j=1}^N |\phi_j(x)|^2,
\een
and the kinetic energy is
\ben
T = \frac{1}{2}\sum_{j=1}^N \int_0^1 dx |\phi'_j (x)|^2.
\een
A dataset is created by randomly sampling $a_j \in [1,10], b_j \in [0.4,0.6], c_j \in [0.03,0.1]$,
to generate 2000 different potentials.
For each potential, the system is occupied with up to 4 fermions, and the exact densities
and kinetic energies are computed. 
Numerically, the Schr\"{o}dinger equation is solved by discretizing the density on a grid:
\ben
x_j = (j - 1)/(N_G - 1), \quad j=1,\dots,N_G
\label{eq:grid}
\een
and $\Delta x=1/(N_G-1)$ is the grid spacing.
Numerov's method~\cite{HNPW93} together with a shooting method is used to solve for the eigenvalues
and eigenfunctions of \Eqref{eq:scheq}. For $N_G=500$, the error in our reference kinetic energies
is less than $10^{-7}$.
\Figref{f:data_example} gives a few sample densities and their corresponding potentials.

The data used here is identical to that of Ref.~\cite{SRHM12}. The exact values of
the parameters used in each sample is given in the supplementary information of 
Ref.~\cite{SRHM12}.
Of the 2000 samples generated, the first half is reserved for training while the second half
is reserved for testing (which we refer to as the {\em test set}).

\ssec{Orbital-free DFT}
\label{sect:ofdft}

In orbital-free DFT, $T\s$ is approximated as a functional of $n(x)$. For our model system with
non-interacting fermions, the total energy is given as
\ben
E_v = \min_n \left\{T[n] + V[n]\right\},
\een
for a given potential $v(x)$. The potential is known exactly as a functional of $n(x)$:
\ben
V[n] = \int_0^1 dx\, n(x) v(x).
\label{eq:potentialenergy}
\een
Via the variational principle, the ground-state
density is found by the Euler-Lagrange constrained search 
\ben
\delta \left\{ E_v[n] - \mu \left( \int n(x)\,dx - N \right) \right\} = 0,
\label{eq:eulereq}
\een
where the chemical potential $\mu$ is adjusted to produce the required particle number $N$.
This becomes simply
\ben
\frac{\delta T[n]}{\delta n(x)} = \mu - v(x).
\label{eq:totEmin}
\een
The density that satisfies this equation, minimizing $E_v[n]$ with the
normalization constraint, is found self consistently.

Given the exact functional $T[n]$, solving Eq.~(\ref{eq:totEmin}) will yield the exact ground-state
density of the system. But in practice, $T$ must be approximated. Let $\tilde T$ be such an approximation, $n(x)$ be the exact density, and $\tilde{n} (x)$ be the self-consistent density found with $\tilde T$.
There are two measures of the error of such an approximate $\tilde T$~\cite{KSB13}.
The first is to compute the functional-driven error $\Delta T_F = \tilde T[n] - T[n]$,
which is simply the error in the KE evaluated on the exact density.  
 The second (and much more difficult) test is to 
insert $\tilde T$ into Eq.~(\ref{eq:totEmin}), solve for the approximate density $\tilde n$, and compute its error relative
to the KE of the exact density $\Delta E = \tilde E_v[\tilde n] - E_v[n]$. Then the density-driven error is defined as
 $\Delta E_D = \Delta E - \Delta T_F$ \cite{KSB13}. This is the additional error incurred by the approximate density.
In practice, a functional which only satisfies the first test is not much use, as the ground-state density itself
must also be obtained from this approximation. In orbital-free DFT, self-consistent
results can be much worse than energies of KS densities, as inaccuracies in the functional derivative can cause large errors
in the corresponding density. In the case of the KE functional for real systems, functional derivatives of
traditional approximations can have singularities at the nuclei, making all-electron calculations very
difficult, if not impossible, to converge~\cite{KT12}. Many of these problems can be avoided through use of pseudopotentials~\cite{KT12,XHSC12},
but in general the solution for Eq.~(\ref{eq:totEmin}) is nontrivial.

As mentioned above, the simplest density functional approximation to $T\s$ is the local approximation~\cite{DG90},
which for spin-polarized densities in 1d is
\ben
T\loc[n] = \frac{\pi^2}{6} \int dx\, n^3(x).
\een
For $N=1$, the exact KE has the von Weizs{\"a}cker~\cite{W35} form:
\ben
T\W[n] = \int dx\, \frac{n'(x)^2}{8n(x)}.
\een
As was shown in Ref.~\cite{SRHM12}, the local approximation does poorly.
The mean absolute error (MAE) on the test set is 217 kcal/mol, and
self-consistent results are even worse at 1903 kcal/mol.
A standard extension of the local approximation to a semi-local form is to add
a fraction of $T\W[n]$ to $T\loc[n]$, forming a modified gradient expansion approximation.
It was shown in Ref.~\cite{SRHM12} that this did little to improve upon the
local approximation.



\ssec{Data topology and representation}

Typically in ML, the data has a finite representation. For example, in \Ref{RTML12}, molecular structures are represented by a Coulomb matrix and the model predicts atomization energies.
In contrast, the electronic density $n(x)$ is a continuous function restricted to the domain~\cite{L02}
\ben
\J_N \equiv \left\{ n \,\Big|\, n(x) \geq 0, n^{1/2}(x) \in H^1(\mathbb{R}), \int n(x)\, dx = N\right\},
\een
where $H^1(\mathbb{R})$ is a Sobolev space\footnote{
A Sobolev space $W^{k,p}(\mathbb{R})$ is a vector space of functions with a norm that is a combination of $L^p$-norms of the function itself and its derivatives up to a given order $k$. It is conventional to write
$W^{1,2}(\mathbb{R})$ as $H^1(\mathbb{R})$. $f \in H^1(\mathbb{R})$ means that $f$ and its first order derivative are in $L^2$.}.
Although $\J_N$ is infinite dimensional,
in practice $n(x)$ is expanded in a finite basis (with $N_G$ basis functions).
In this work, we use a real space grid to represent the density, since our reference calculations
are done using the same grid.
We use the $L^2$ inner product and norm between densities
$n_i(x),n_j(x)$
\ben
\langle n_i, n_j \rangle = \int_{-\infty}^\infty dx\, n_i(x) n_j(x), \quad \| n \| = \sqrt{\langle n, n \rangle}.
\label{eq:innerprod}
\een
(In actual calculations, all densities are represented on a
finite basis, and thus will have have a finite $L^2$-norm).
Since the ML algorithm is expressed in terms of this inner product,
the results are independent of the specific representation used
as long as the basis is converged.

Even with a truncated basis, $\J_N$ is still high-dimensional and applying ML to learn
the KE of all densities in $\J_N$ would not be feasible. Fortunately, we are only interested in a subspace of $\J_N$ related to a specific
class of potentials (e.g. Gaussian dips), which greatly reduces the 
variety of possible densities. 
In general, let the potential $v(x)$ be parametrized by the parameters $\{p_1, \dots, p_d\}$.
We define the density manifold $\M_N \subset \J_N$ as the set of all densities that come from 
these potentials with a given particle number $N$. In general, $\M_N$ is a $d$-dimensional manifold.
The training densities, $n_j(x)$ for $j=1,\dots,N_T$, are sampled from $\M_N$.
In the present work, the external potential has 9 parameters, and thus $d$ is at most 9.

\ssec{The kernel trick and feature space}

In finding the structure of low-dimensional data, it is often sufficient to optimize parametrized 
non-linear forms (e.g., using a polynomial to fit a sinusoid). For high-dimensional, nonlinear data
this becomes increasingly difficult. In kernel-based machine learning, the approach is to transform the 
{\em data itself} non-linearly to a high-dimensional space known as feature space, 
such that the data becomes linear \cite{V95,SMBK99,MMRT01,MBKM13,SS02}.
\begin{figure}[tb]
\includegraphics[width=\columnwidth]{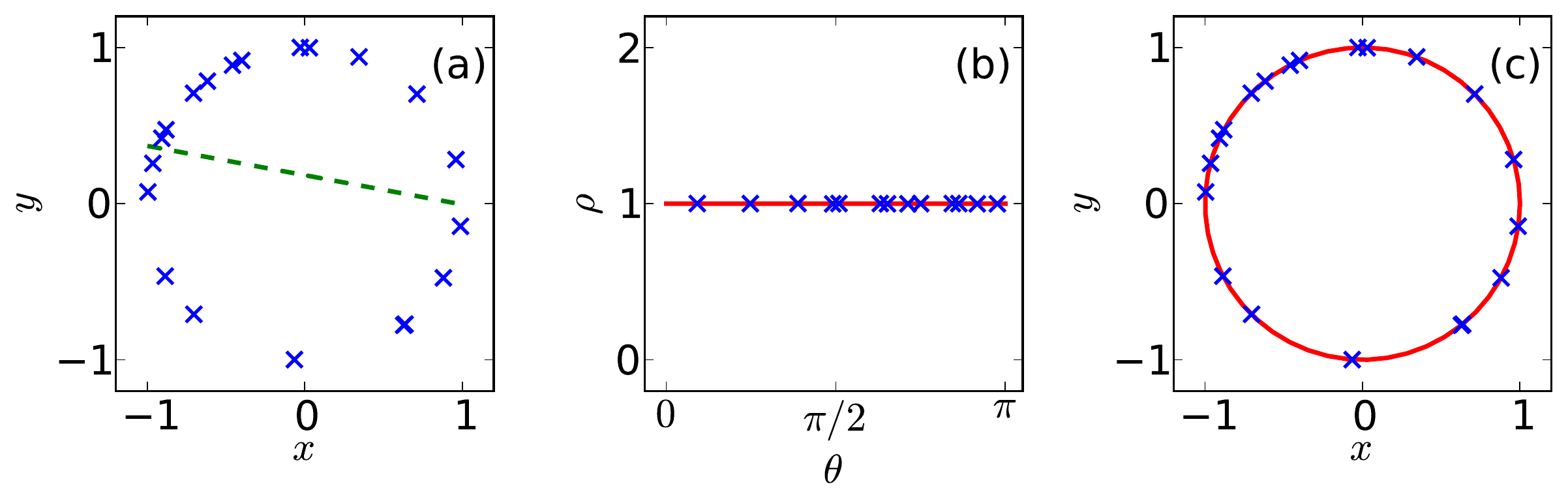}
\caption{Example of the non-linear transformation of data to feature space. (a) The data is non-linear (a circle) in Cartesian coordinates. The green dashed line is a linear fit to the data points (blue crosses).
(b) When the data is transformed to feature space by $x\to \rho \cos \theta$, $y \to \rho \sin \theta$, the linear
structure in the data is revealed (red solid line). (c) The model can be transformed back to the original space
to give a non-linear fit of the data.}\label{f:nonlinear}
\end{figure}

\Figref{f:nonlinear} illustrates data points that lie on a circle in the Cartesian plane. 
As shown, the data becomes linear on transformation to polar coordinates, and linear
regression can subsequently be used to fit the data. 
Transforming back to Cartesian coordinates recovers the non-linearity.
Let the data points belong to a vector space $\chi$, also called input space, and let
$\Phi : \chi \to F$ be the map to feature space $F$. Assuming we wish to apply 
a linear method such
as regression in feature space $F$, we note that regression can be expressed solely in
terms of the inner product between feature vectors $\Phi(x)$ and $\Phi(y)$, where $x,y\in \chi$. 
We define the kernel $k$ such that
\ben
k(x,y) = \langle \Phi(x), \Phi(y) \rangle.
\label{eq:kerneldef}
\een
The kernel can generally be thought of a measure of similarity 
between data, but must satisfy Mercer's condition:
\ben
\int\int k(x,y) g(x) g(y) dx dy \geq 0,
\label{eq:Mercercondition}
\een
for all $g(x)$ satisfying $\int^{\infty}_{-\infty}|g(x)|^2 dx<\infty$.
Mercer's theorem~\cite{M09} guarantees the existence
of a feature space $F$, which is a reproducing kernel Hilbert space~\cite{A50}.
Since the linear algorithm in $F$ may be expressed in terms of the kernel 
in \Eqref{eq:kerneldef}, $\Phi$ need never be explicitly computed. 
This procedure is known as the {\em kernel trick}, and enables easy nonlinearization
of all linear scalar product-based methods that can be expressed via an inner product \cite{SSM98}.

\ssec{Kernel ridge regression}

Kernel ridge regression is a nonlinear version of regression with a regularization term to prevent overfitting~\cite{HTF09}. Our MLA for the KE has the form
\ben
T\ML[n] = \sum_{j=1}^{N_T} \alpha_j k[n, n_j],
\label{eq:TMLdef}
\een
where $N_T$ is the number of training densities, $\alpha_j$ are weights to be determined, $n_j$ are 
training densities and $k[n, n_j]$ is the kernel.
The weights are found by minimizing the quadratic cost plus regularization
\ben
{\mathcal C}(\ba) =  \sum_{j=1}^M (T\ML[n_j] - T[n_j])^2 +  \lambda \ba^\top \K \ba,
\label{eq:costfunction}
\een
where $\ba = (\alpha_1, \dots, \alpha_{N_T})$,
$\K$ is the kernel matrix, $\K_{ij} = k[n_i, n_j]$, and $\lambda$ is called the 
regularization strength. The second term penalizes weights with large magnitudes in order to prevent overfitting.\footnote{
The regularization term accounts for the possibility of noisy data (e.g. experimental data),
and imposes certain smoothness conditions on the model (see \cite{SmoSchMue98}).  
Our reference data is deterministic and thus noise-free in this sense,
but, because the precision of our calculations is limited, we may consider the
numerical uncertainty to be noise.
}
By setting the gradient of Eq.\ref{eq:costfunction} to zero, minimizing ${\mathcal C}(\ba)$ gives
\ben
\ba = (\K + \lambda \matrixsym{I})^{-1} \Tv,
\label{eq:KRRsolution}
\een
where $\matrixsym{I}$ is the identity matrix and $\Tv~=~(T[n_1], \dots, T[n_{N_T}])$.
The {\em hyperparameters}, which include the regularization strength $\lambda$ 
and the parameters of the kernel such as the length scale $\sigma$, are found via cross 
validation (see \cite{HMBF13} and Sect. \ref{MODELSELECTION}).

\begin{figure}[tb]
\centering
\includegraphics[width=\figwidth]{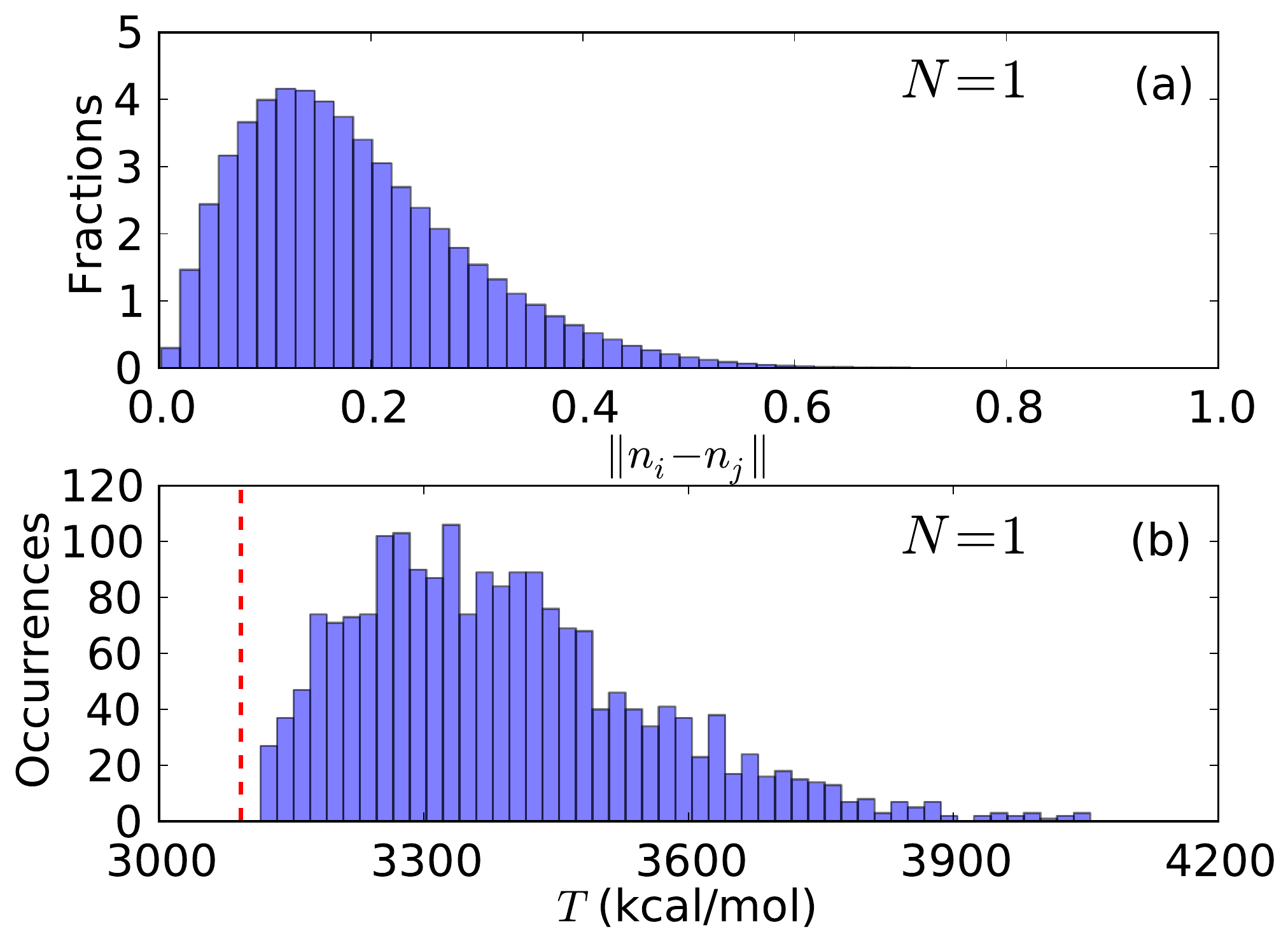}
\caption{(a) Normalized distribution of the Euclidean distance between all distinct pairs of densities in the dataset (2000 densities). The maximum distance between any pair is $0.9$. (b) Histogram of the KE in the dataset. The vertical dashed line at 3093 kcal/mol is the ground-state energy of one fermion in a flat box of length 1.}
\label{f:NT_2000_distri}
\end{figure}

The choice of the kernel will depend on the given data.  Some kernels
are designed to be generally robust and applicable (e.g., the Gaussian
kernel), while others are designed for a specific type of data (see
e.g.~\cite{RPS07,ZRMS00,MMRT01}). A good choice of kernel can reflect the
characteristics of the data (see \cite{BBM08}).  In~\Ref{SRHM12}, we
chose the Gaussian kernel
\ben
k[n_i, n_j] = \exp\left(-\| n_i - n_j \|^2/2\sigma^2 \right),
\label{eq:gaussiankernel}
\een
where $\sigma$ is the length scale. 
Since the density is represented on a uniform grid, the $L^2$-norm can be approximated
by\footnote{
Note that, in~\Ref{SRHM12}, the same representation for the density was used, but the 
density were treated as vectors, so the standard Euclidean distance was used in the kernel. This is equivalent
to the formulation here, except our notation is more general now (e.g. Simpson's rule could be used to
approximation the $L^2$-norm instead of a Riemann sum), and the length scale in Gaussian kernel here is related to the
scale of the kernel in~\Ref{SRHM12} by a factor of $\sqrt{\Delta x}$.
}
\ben
\| n_i - n_j \|^2 = \Delta x \sum_{l=1}^{N_G} (n_i(x_l) - n_j(x_l))^2
\label{eq:density_distance}
\een
where $x_l$ is given by the grid defined in \Eqref{eq:grid}.
This approximation becomes exact as $\Delta x \to 0$.
\Figref{f:NT_2000_distri} shows the range and distribution of Euclidean distances between all pairs of densities and KE of all densities in the dataset with $N=1$. 

\begin{figure}[tb]
\centering
\includegraphics[width=\columnwidth]{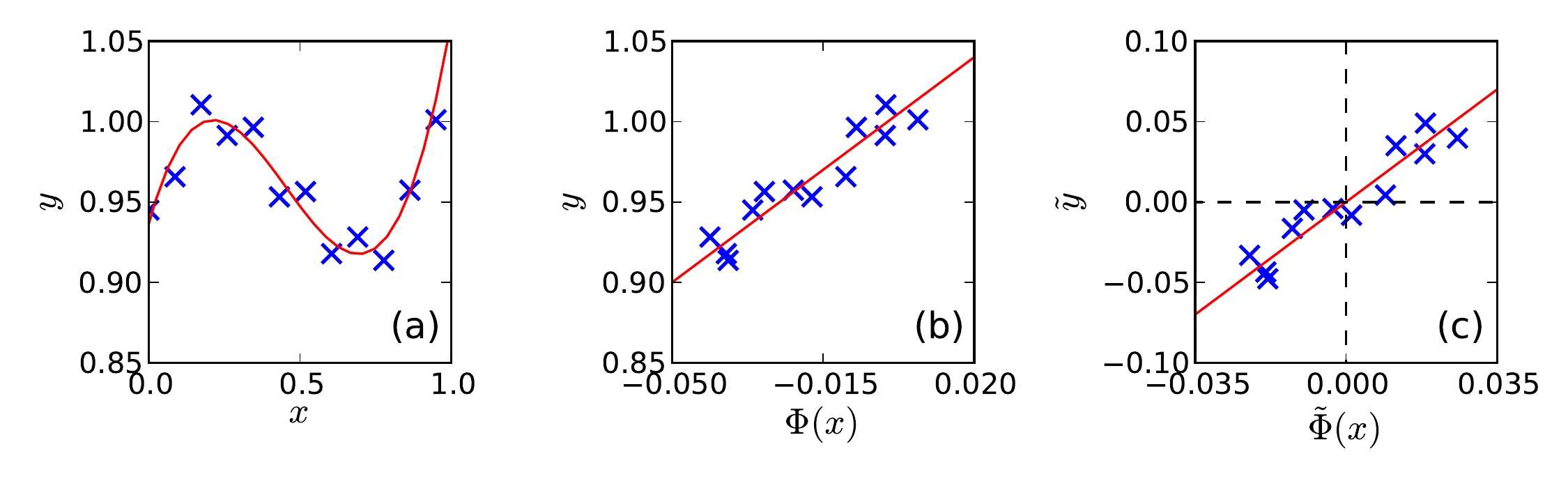}
\caption{(a) An example 1d noisy data set. (b) Transformation to feature space $\Phi(x)$. (c) Centering of data in feature space.}
\label{f:feature}
\end{figure}

\begin{figure*}[tb]
\includegraphics[width=17cm]{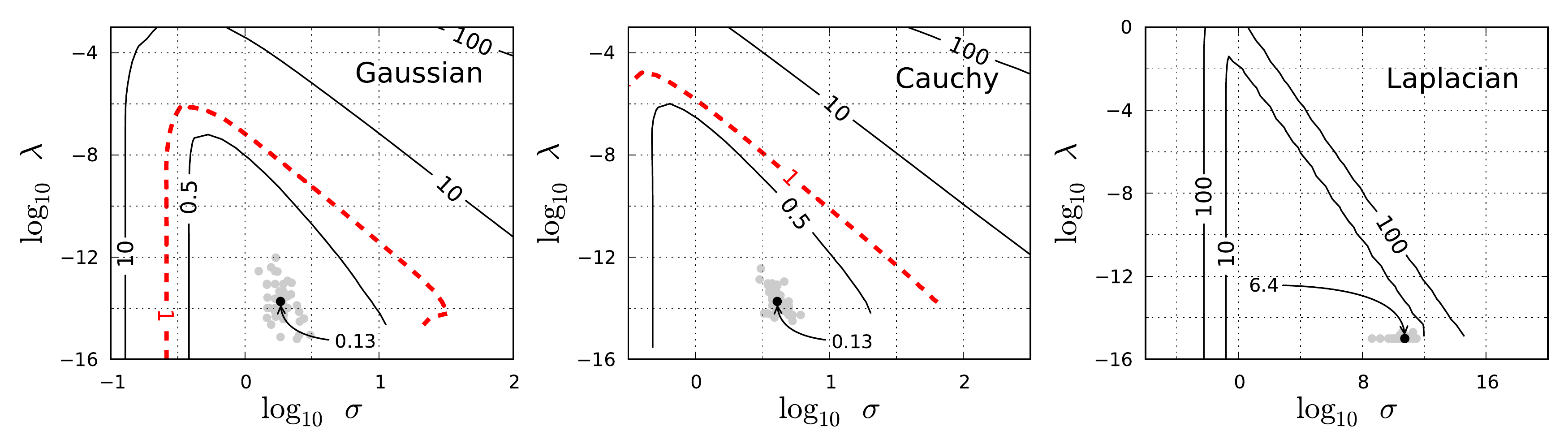}
\caption{Contour plots of the functional-driven MAE $\overline{|\Delta T_F|}$ over the test set
in kcal/mol for selected kernels with $N_T=100$.
The dashed line delineates the region where the model achieves chemical accuracy.
Each gray dot gives the optimal choice of hyperparameters from a randomized 
10-fold cross validation. The black dot denotes the median over 40 repetitions.
In the lower right region (i.e. small $\lambda$ and large $\sigma$), the matrix inverse in \Eqref{eq:KRRsolution} is numerically unstable due to the limited precision of the calculation.
}
\label{f:cv_surface}
\end{figure*}
Ordinary linear regression models frequently employ a bias term to account for the fact that the data might lie away from the origin.
Without this term, the regression line is forced to go through the origin, causing a systematic error if the  data does not.
The bias term can be implemented directly, or by centering the samples such that the mean is zero.
\Figref{f:feature} illustrates the transformation to feature space for an example 1d data set and
linear regression in feature space. If the data is centered in feature space, the bias term is unnecessary.
Here, we center the densities in features space such that $\sum_{j=1}^{N_T}\Phi(n_j)~=~0$. We define the centered map to feature space
$\tilde \Phi(n) = \Phi(n) - \sum_{j=1}^{N_T} \Phi(n_j)/N_T$.
Then the centered kernel is \cite{SSM98}
\bea
\tilde k[n, n'] &=& \langle \tilde\Phi(n), \tilde\Phi(n') \rangle \nonumber \\
	&=& k[n, n'] - \frac{1}{N_T} \sum_j^{N_T} (k[n', n_j] + k[n, n_j]) \nonumber \\
	&& {} + \frac{1}{N_T^2} \sum_{i,j=1}^{N_T} k[n_i, n_j].
\eea
For simplicity, all equations given in this work assume that the data is centered (i.e. $k=\tilde k$).
In fact, kernels such as the Gaussian kernel Eq.(\ref{eq:gaussiankernel}) 
whose induced reproducing kernel Hilbert space
on a bounded domain is dense in the space of continuous functions on this 
domain do not require centering~\cite{PMRR01}.


\sec{Model selection}\label{MODELSELECTION}

\ssec{Kernels}

Model selection refers to the process of selecting a kernel and the corresponding
hyperparameters. In kernel ridge regression, this includes the regularization strength
$\lambda$ and the kernel parameters (e.g. in the Gaussian kernel, the length scale $\sigma$).
Table~\ref{tbl:kernelforms} lists some standard kernels.
\renewcommand*\arraystretch{1.4}
\begin{table}[tb]
\begin {tabular}{|@{\hspace{1.0em}}c@{\hspace{1.0em}}|@{\hspace{1.0em}}l@{\hspace{1.0em}}|}%
\toprule [0.05em]
Kernel & $k[n, n']$ \\
\midrule [0.01em]
Gaussian & $\exp( -\| n - n' \|^2/2\sigma^2 )$ \\
Cauchy & $(1+\| n - n' \|^2/\sigma^2)^{-1}$ \\
Laplacian & $\exp(-\| n - n' \|/2\sigma )$ \\
Wave & $\displaystyle\frac{\theta}{\| n - n' \|} \sin \frac{\| n - n' \|}{\theta}$ \\
Power & $\|n - n' \|^d$ \\
Linear & $\langle n, n' \rangle$ \\
\bottomrule [0.05em]%
\end{tabular}
\caption{Standard kernels. The parameters $\sigma$, $\theta$, $d$ are kernel parameters. 
The linear kernel has no parameters.}
\label{tbl:kernelforms}
\end{table}
Radial basis function (RBF) kernels, which include the Gaussian, Cauchy, and Laplacian kernels,
all behave similarly and tend to work for a broad range of problems. 
Other kernels work well for specific data structures~\cite{RPS07,MMRT01,ZRMS00} and 
regularization properties~\cite{SSM98}.

\Figref{f:cv_surface} shows the contours of the functional-driven MAE
over the test set as a function of the 
regularization strength $\lambda$ and the kernel
parameter $\sigma$. We see that the qualitative behavior is similar
for the Gaussian, Cauchy and Laplacian kernels. 
In the left region (where
the contour lines are vertical), the length scale $\sigma$ is much smaller than the distance between
neighboring training densities. Thus the RBF-type kernel functions centered at each training density have
minimal overlap, yielding a poor approximation to the KE functional. 
The kernel matrix becomes nearly unity, and the regularization $\lambda$ has negligible effect.
On the right side of the contour plot, the length scale is comparable to the global scale of the data. 
In these regions, the kernel functions are slowly varying and do not 
have enough flexibility to fit the nonlinearity in the data.
The region with minimum MAE lies in the middle.
The Gaussian and Cauchy kernels both give the same performance, with errors less than 1 
kcal/mol in the middle region (enclosed by the dashed line), while the Laplacian kernel 
behaves poorly in comparison.
This is likely due to the cusp in the form of the kernel, which cannot fit the smooth
KE functional.

\ssec{Optimization of hyperparameters}

After picking a kernel family, the values of the hyperparameters must be chosen.
Ideally, we select the hyperparameters such that the
generalization error, which is the error not only on our training set but
also on all future data,
is minimal. The out-of-sample error must be estimated without looking at the test set (the test set
is never touched during model selection, so that it can give a true test
of the final performance of the model) \cite{HMBF13,MMRT01}. 
This procedure, known as cross-validation, is essential for model selection
in preventing
overoptimistic performance estimates (overfitting). 

\begin{figure}[tb]
\centering
\includegraphics[width=\figwidth]{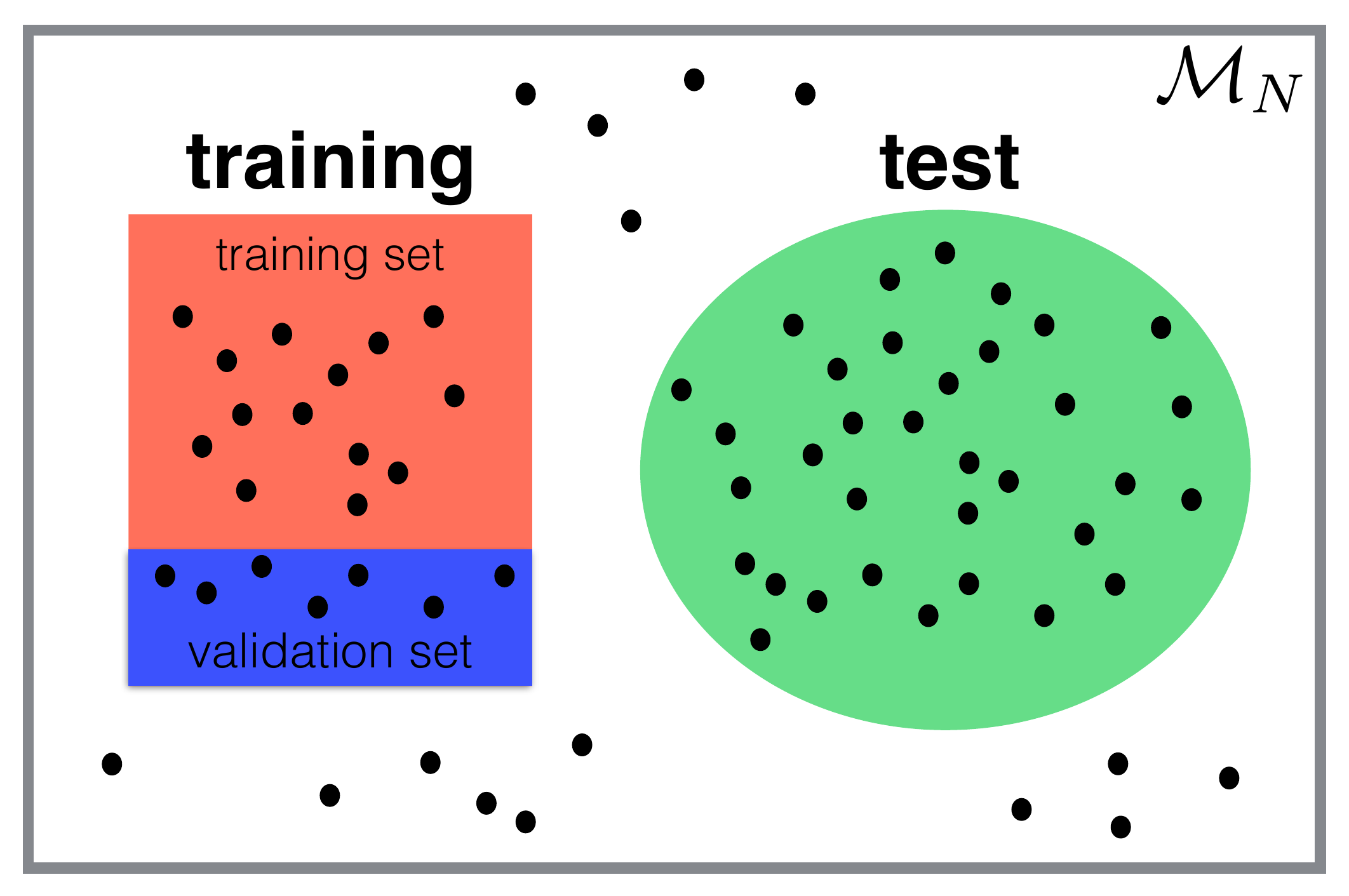}
\caption{Cartoon shows the relation of each data set in $\M_N$. Each black dot represents a sample (density and its corresponding KE). Training, validation and test set are subsets of the full data set.}
\label{f:cvcartoon}
\end{figure}

Various schemes for cross validation exist \cite{HMBF13,MMRT01,AMMF97},
but all obey a basic principle: the available data is subdivided into
three parts: the {\em training}, {\em validation} and the {\em test
sets}. 
The ML model is built from the training set and the hyperparameters
are optimized by minimizing the error on the validation set
(\Figref{f:cvcartoon}).  The test set is never touched until the
 weights and hyperparameters
have been determined.
Then and only then, the generalization ability of the model can be
assessed with the test data (\cite{MMRT01,HMBF13}, 
see also \cite{lemm2011introduction}).
Typically, the data is shuffled to ensure its random distribution between training and validation division.
This can be repeated with different subdivisions. A few schemes, which will be analyzed for our 
kinetic energy functional estimation problem, are described below. For each scheme,
a test set of 1000 samples is used to estimate the generalization error after the ML model
is selected.

\begin{description}
\item[Simple cross validation]
The training data ($N_T$ samples) is randomly divided into a training set of $70\%$ 
and a validation set of $30\%$. The hyperparameters are optimized by minimizing 
the MAE on the validation set.

\item[$k$-fold cross validation]~

\begin{description}
\item[Step 1] The $N_T$ training data is randomly divided into $k$ bins.

\item[Step 2] The $j$th bin is used as the validation set and the remaining $k-1$ bins as training set.
The model is built on the training set and the hyperparameters are selected by
minimizing the MAE on the validation set.

\item[Step 3] Repeat step 2 $k$ times such that all bins have been used as validation sets. 
We will then have $k$ models in
total and the final hyperparameters are selected as the median over all models.
\end{description}
Because the mean cross validation error still depends on the initial random partitioning of data in cross validation, we repeat the procedure
with different subdivisions \cite{MMRT01}.

\item[Leave-one-out]
Leave-one-out (LOO) is a special case of $k$-fold cross validation, when $k=N_T$. Thus each bin contains only one sample.
\end{description}

Typically, it is better to leave out as little data as possible to exploit the statistical power in the data.
Simple cross validation is computationally expedient, but wasteful since not all training data 
participates in the optimization.
$k$-fold cross validations are used
in situations where data is very limited, or expensive to collect.
Leave-one-out is often used with limited data and it becomes computationally intensive if $N_T$ is large. 
$k$-fold cross validation gives a good balance on all counts.


\begin{figure}[tb]
\centering
\includegraphics[width=\figwidth]{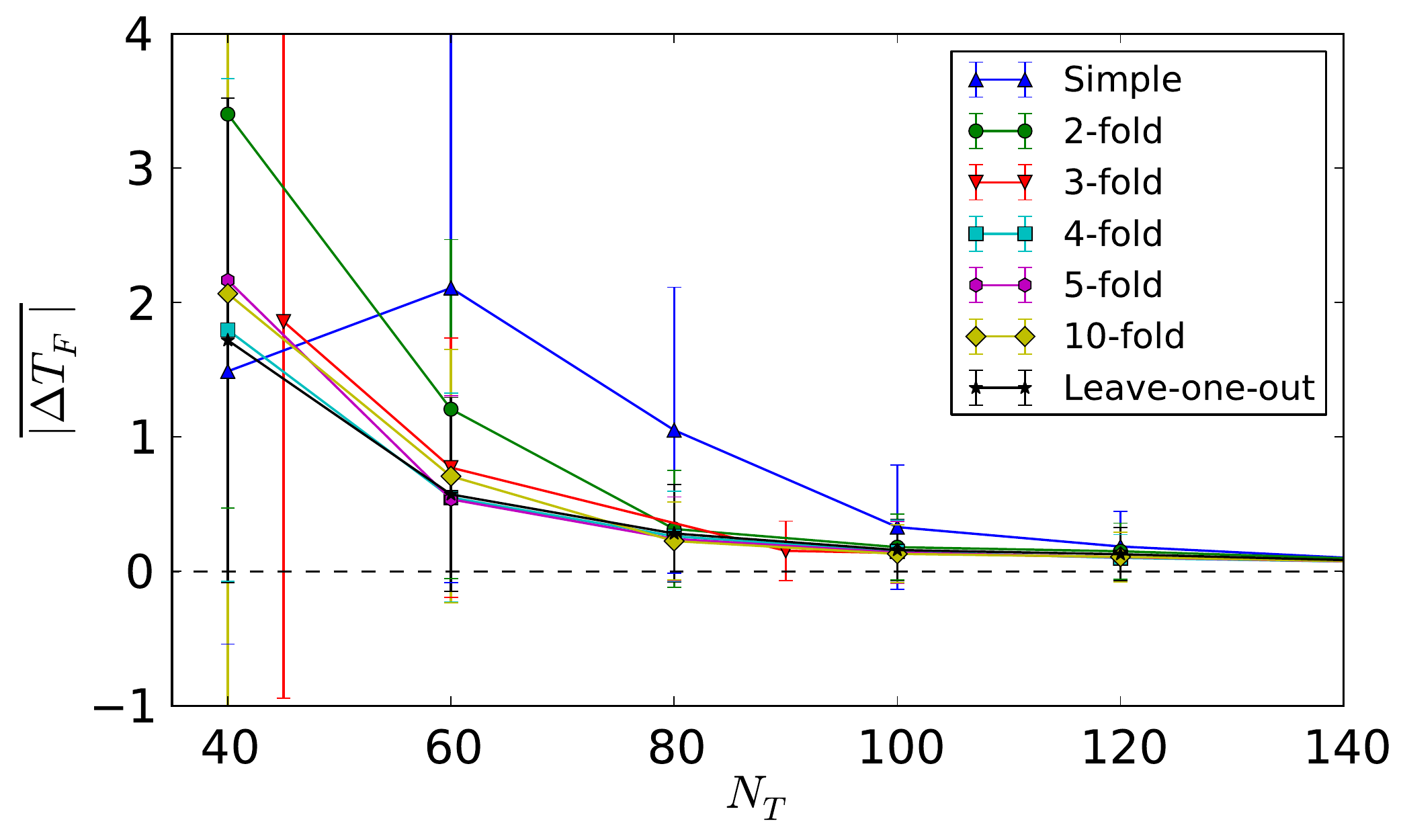}
\caption{Comparison of different cross validation methods, including simple, 2-fold, 3-fold, 4-fold, 5-fold, 10-fold and leave-one-out. The mean of the absolute functional-driven error $|\Delta T_F|$ (in kcal/mol) is evaluated on the test set and the error bars represent the standard deviation.}
\label{f:diff:cv}
\end{figure}


For the Gaussian kernel, 
\Figref{f:diff:cv} shows the MAE on the test set with the hyperparameters optimized with different cross validation methods. With 120 training densities, all schemes give a similar MAE,
despite the large variations in $\sigma$ and $\lambda$. This means that multiple
models exist that give comparable performance. As expected, the only variations in MAE
occur for more limited data.

\Figref{f:cv_surface} shows how $10$-fold cross validation performs in selecting hyperparameters that generalize well to the test set, for a few kernels.
The gray dots represent the optimal parameter choice for each repetition, and
the black dot is the median over all repetitions. In this case, the global minimum of the MAE
lies in a relatively flat basin. Each randomized cross validation lies near the true minimum,
indicating the model generalizes well to the test set.

\renewcommand*\arraystretch{1.4}
\begin{table}[t]
\begin {tabular}{|l|c|c|c|c|}
\toprule [0.05em]
Kernel & $\lambda$ & $p$ & $\overline{|\Delta T_{F}|}$ & $|\Delta T_{F}|^{\text{max}}$  \\
\midrule [0.01em]%
Gaussian		& $4.5\cdot 10^{-14}$	& 1.6		& 0.13 & 3.4 \\
Cauchy		& $7.8\cdot 10^{-14}$	& 3.5		& 0.13 & 2.9 \\
Laplacian	& $1.0\cdot 10^{-15}$	& $3.6  \cdot 10^5$		& 6.4 & 231 \\
Linear		& $6.2\cdot 10^{-1}$		& -		& 53.1 & 380 \\
Wave		& $4.5\cdot 10^{-1}$		& 0.14		& 19.2 & 252 \\
Power		& $1.0\cdot 10^{-13}$		& 1.96		& 3.3 & 104 \\
\bottomrule [0.05em]%
\end {tabular}%
\caption{The optimal hyperparameters found through 10-fold cross validation
 and the MAE over the test set for 
various kernels with $N=1$ and $N_T=100$. The kernel parameter 
$p$ refers to $\sigma$ for the Gaussian, Cauchy 
 and Laplacian kernels, $\theta$ for the wave kernel and $d$ for the power kernel. The linear
kernel has no parameter. Errors are given in kcal/mol.
}\label{t:ML:Tperf:short2}
\end{table}
Finally, we use 10-fold cross validation (repeated 40 times) to optimize the hyperparameters.
\Tabref{t:ML:Tperf:short2} shows the optimal hyperparameters and functional driven errors for 
the kernels listed in~\Tabref{tbl:kernelforms}.
Some optimum values for the Gaussian kernel are listed in~\Tabref{t:ML:Tperf:short}.
Detailed information with optimum values of other kernels are shown in
supplementary material.

\sec{Results and Discussion}

In the main work of this paper, we test in greater detail some of the
methods that were introduced in~\Ref{SRHM12} using only the Gaussian kernel,
as it performs the best.

\renewcommand*\arraystretch{1.4}
\begin{table*}[t]
\begin {tabular}{|D{.}{.}{3.1}|D{.}{.}{4.1}|D{.}{.}{4.3}|D{.}{.}{3.4}|D{.}{.}{3.5}|D{.}{.}{3.3}|D{.}{.}{3.5}|D{.}{.}{5.3}|D{.}{.}{3.5}|D{.}{.}{4.3}|}

\toprule [0.05em]

\multicolumn{1}{|c|}{} & \multicolumn{1}{c}{} & \multicolumn{2}{|c}{} & \multicolumn {2}{|c|}{$|\Delta T_{F}|$} & \multicolumn {2}{c}{$|\Delta T |$} &
\multicolumn {2}{|c|}{$|\Delta E|$} \\

\multicolumn {1}{|c|}{$N$}		& \multicolumn {1}{c}{$N_T$}	& \multicolumn {1}{|c}{$\lambda\cdot 10^{14}$}	& 
\multicolumn {1}{c}{$\sigma$}	& \multicolumn {1}{|c}{Mean}	& \multicolumn {1}{c|}{Max}		&
\multicolumn {1}{c}{Mean}		& \multicolumn {1}{c}{Max}	& \multicolumn {1}{|c}{Mean}		&
\multicolumn {1}{c|}{Max} \\

\midrule [0.01em]%

1		& 40		& 50. 	& 4.2	& 1.9		& 30. 	& 15 	& 120 	& 5.1 	& 32 \\
		& 60		& 10. 	& 1.8	& 0.62		& 11. 	& 3.0 	& 19 	& 0.66 	& 4.4 \\
		& 80		& 54.	& 1.5	& 0.23		& 3.1 	& 1.1 	& 11 	& 0.44 	& 2.6 \\
		& 100		& 4.5	& 1.6	& 0.13		& 3.5 	& 1.4 	& 16 	& 0.41 	& 2.3 \\
       & 150    	& 1.2   & 1.3   & 0.06 		& 1.0 	& 0.81 	& 5.1	& 0.27	& 1.9 \\
      	& 200    	& 1.3   & 1.0   & 0.03 		& 0.87 	& 0.67 	& 10.	& 0.28	& 1.6 \\
\midrule [0.005em]%
2      	& 60     	& 60.	& 3.0   & 0.46 		& 4.8 & 1.79 	& 9.9 	& 0.73 	& 3.6 \\
      	& 100    	& 1.0	& 2.2	& 0.14 		& 1.7 & 1.25 	& 5.0 	& 0.44 	& 2.5 \\
\midrule [0.005em]%
3      	& 60     	& 6.0	& 5.8	& 0.31		& 3.9 & 1.03 	& 5.0 	& 0.82 	& 6.5 \\
      	& 100    	& 1.9	& 2.5	& 0.13		& 1.7 & 1.11 	& 8.3 	& 0.59 	& 3.8 \\
\midrule [0.005em]%
4      	& 60     	& 0.6	& 14	& 0.46		& 5.4 & 2.44 	& 9.5 	& 0.93 	& 6.3 \\
      	& 100    	& 1.4	& 2.7	& 0.08		& 2.6 & 1.12 	& 9.8 	& 0.63 	& 5.0 \\
\midrule [0.005em]%
$1-4$	& 400		& 1.7			& 2.2	& 0.12		& 3.0 & 1.28 & 12.6 & 0.52 & 5.1 \\

\bottomrule [0.05em]%
\end {tabular}%
\caption{Hyperparameters and errors measured over the test set using the Gaussian kernel, for different $N$ and $N_T$. 
The regularization strength $\lambda$ and length
scale of the Gaussian kernel $\sigma$ is optimized with $10$-fold cross validation.
The functional-driven error $\Delta T_F = T\ML[n] - T[n]$ is evaluated on
the test set. Mean and max absolute errors are given in kcal/mol.
$\Delta T = T\ML[\tilde n] - T[n]$, gives the error in the KE evaluated on constrained optimal densities.
Likewise $\Delta E = E\ML[\tilde n] - E[n]$.
}\label{t:ML:Tperf:short}
\end{table*}

\ssec{Errors on exact densities}

In \Tabref{t:ML:Tperf:short}, we evaluate our MLA, constructed using the first $N_T$ training
densities in our data set, on the exact densities of the test set and compute
the errors $\Delta T_F = T\ML[n] - T[n]$.
The Gaussian and Cauchy kernels give the best performance.
For the Gaussian kernel with $N=1$ chemical accuracy is achieved
 (i.e. MAE less than 1 kcal/mol) at $N_T=60$. 
Just as we saw in Ref.~\cite{SRHM12}, the performance is systematically improvable with 
increasing number of training densities.
The Laplacian kernel 
gives a mean absolute error (MAE) of $6.9$ kcal/mol at $N_T=100$ (still better than LDA),
which improves as $N_T$ increases.
On the other hand, the performance of the wave kernel
 does not improve as $N_T$ increases (see supplemental information). This indicates the 
form of the wave kernel is not flexible enough to fit the form of the KE functional.

\ssec{Sparse grid}

Note that the choice of $N_G$ used in the reference calculations is needed to converge our
reference energies and densities, but may be larger then the grid needed
to ``converge'' our ML functional. As the ML model depends only on the inner product between densities,
this will typically converge much faster than, e.g. Numerov's method.
To demonstrate this, we define a ``sparse'' grid, $\{x_{s(j-1) + 1} | j=1,\dots,N_G/s \}$, using every
 $s$th point in the grid (we only choose $s$ such that $N_G$ is divisible by $s$).

\begin{figure}[tb]
\includegraphics[width=8cm]{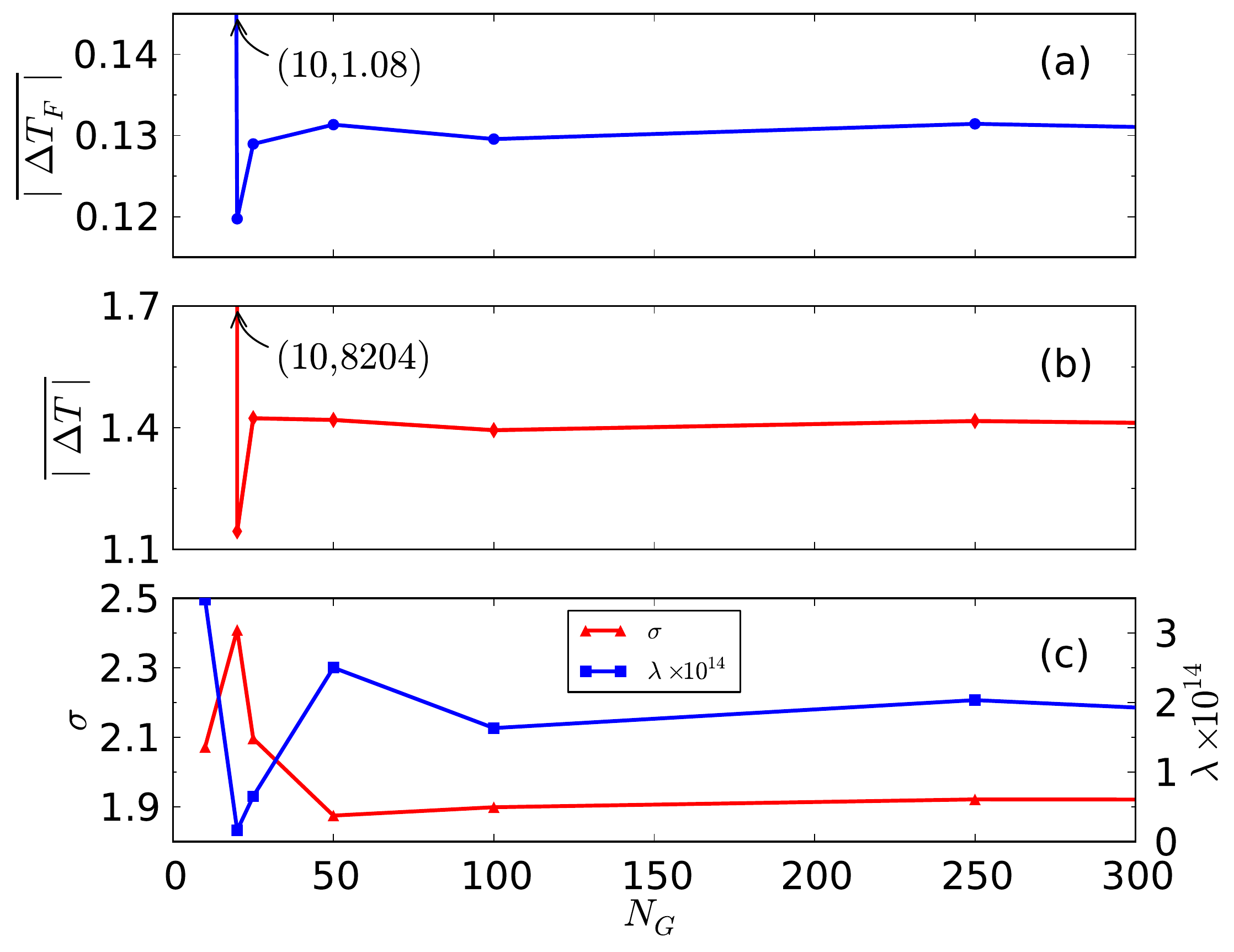}
\caption{The effect of using a sparse grid to represent the density
on the performance of the MLA, for $N=1$, $N_T=100$, with the Gaussian
kernel. Here (a) $\overline{|\Delta T_F|}=\overline{T\ML[n]-T[n]}$ is the mean absolute functional-driven error of the MLA evaluated on the test set in kcal/mol, (b) $\overline{|\Delta T|}=\overline{T\ML[\tilde n]-T[n]}$ gives the error of KE evaluated on constrained optimal densities in kcal/mol and (c) the corresponding re-cross validated hyperparameters $\lambda$ and $\sigma$. The MAE is completely unaffected as $N_G$ is reduced until 
approximately $N_G=10$, when it jumps sharply.}
\label{fig:sparse-grid-N1}
\end{figure}

\Figref{fig:sparse-grid-N1} shows that performance of the model is unaffected until $N_G$ is
reduced to about 10 grid points. The model is cross-validated each time, but the hyperparameters
change only slightly. Thus, ML can accurately learn the KE functional with a far less
complete basis than is required to accurately solve the Schr\"odinger equation. This is possible
because we have restricted the learning problem to a simple type of potential with a limited
range of possible densities and energies. The underlying dimensionality of the data is about 9,
comparable to the number of parameters that determine the potential. 
The model needs only enough degrees of freedom in the representation of the density to distinguish
between densities, but no more. Thus it is no coincidence that the minimum grid required
is comparable to the dimensionality of the data (i.e. the dimensionality of the density
manifold $\M_N$).

However, we also need a sufficiently fine grid to compute the integral in Eq.~(\ref{eq:potentialenergy})
to the desired accuracy. In the problem shown here,
the dimensionality of the data is relatively small, and will increase for larger systems (e.g. real molecules
with many degrees of freedom). In general, however, we need to consider both factors
in choosing a suitable basis. But, we may be able to use a basis that is more sparse
than that of the reference data, which would greatly reduce the computational cost of the method.

\ssec{Challenge of finding density}

\begin{figure}[tb]
\centering
(a)\includegraphics[width=\figwidth]{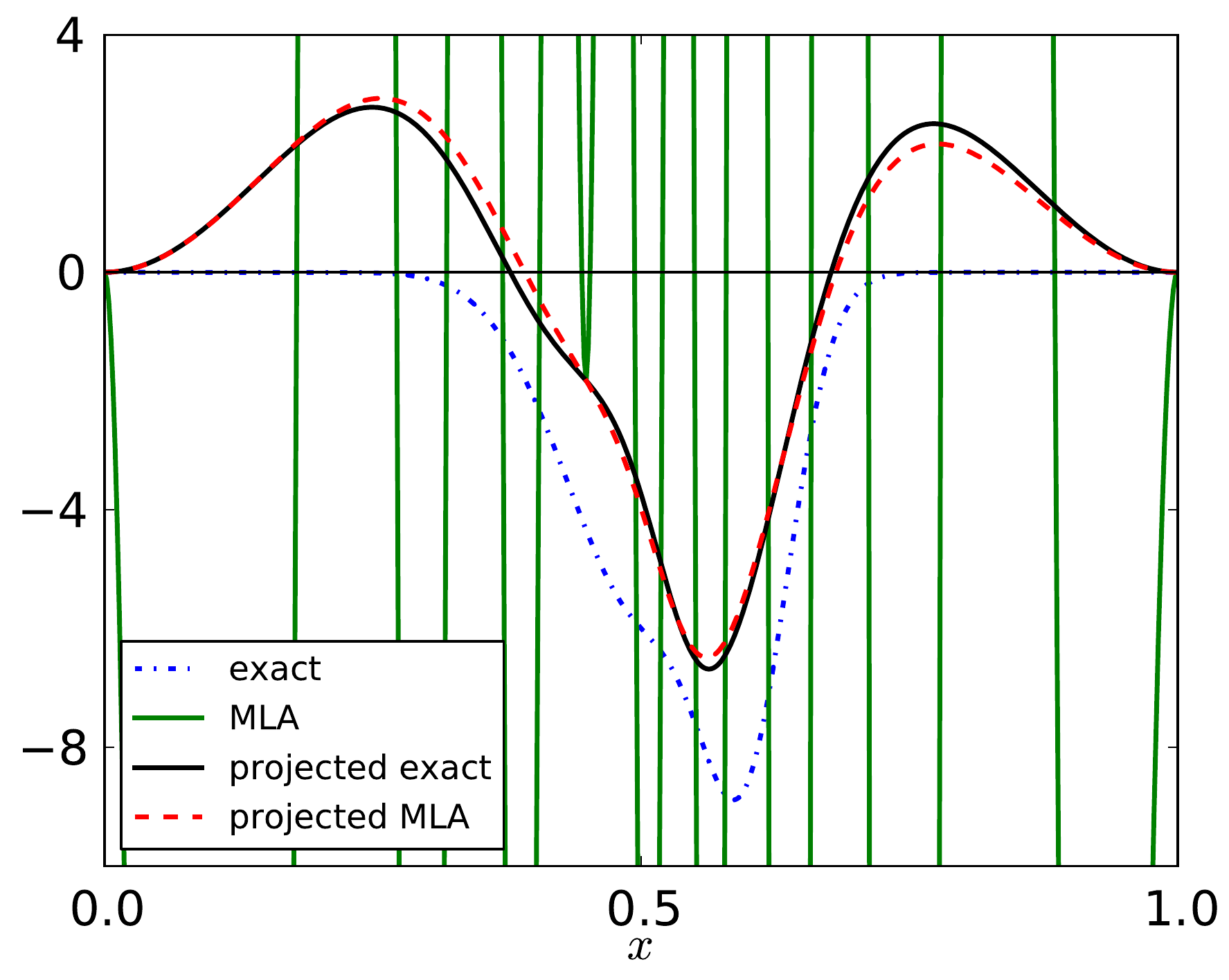}
(b)\includegraphics[width=\figwidth]{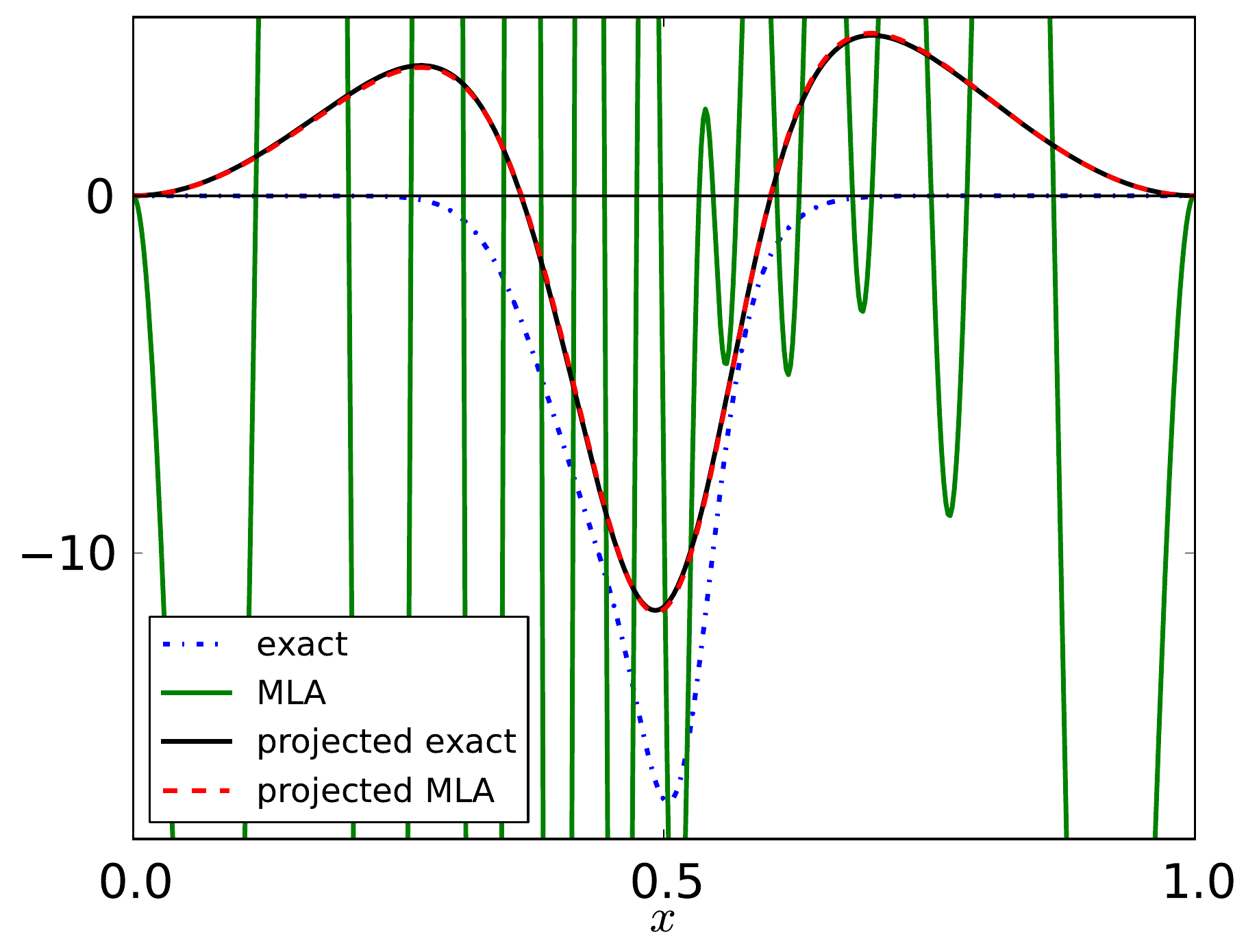}
\caption{The functional derivative of our MLA (green) cannot reproduce the exact
derivative $v(x)$ (blue dot dashed) evaluated at the ground-state density, because
this information is not contained in the data. However, both agree when projected
onto the tangent of the data manifold $\M_N$ at $n$ (black and red dashed). 
Shown for $N=1$, for (a) $N_T=40$ and (b) $N_T=100$, for a typical test sample. }
\label{fig:funcderiv}
\end{figure}

Thus far, we have focused on the discussion of the performance of the MLA evaluated on exact densities
(i.e. the functional-driven errors).
However, in order for a functional to be useful, it must also predict the ground-state density. As discussed previously,
an accurate functional derivative is necessary in order to solve \Eqref{eq:totEmin} and yield
an accurate density. 
The functional derivative of our MLA is given by:
\ben
\frac{\delta T\ML[n]}{\delta n(x)} = \sum_{j=1}^{N_T} \alpha_j \frac{\delta k[n, n_j]}{\delta n(x)},
\een
where, for the Gaussian kernel,
\ben
\delta k[n, n_j]/\delta n(x) = (n_j(x) - n(x)) k[n,n_j] / \sigma^2.
\een
In \Figref{fig:funcderiv}, we plot the functional derivative
of our model compared with the exact derivative. The model displays a highly inaccurate functional derivative,
with a huge amount of apparent ``noise'', as was
found in \Ref{SRHM12}.

\begin{figure}[tb]
\centering
\includegraphics[width=\figwidth]{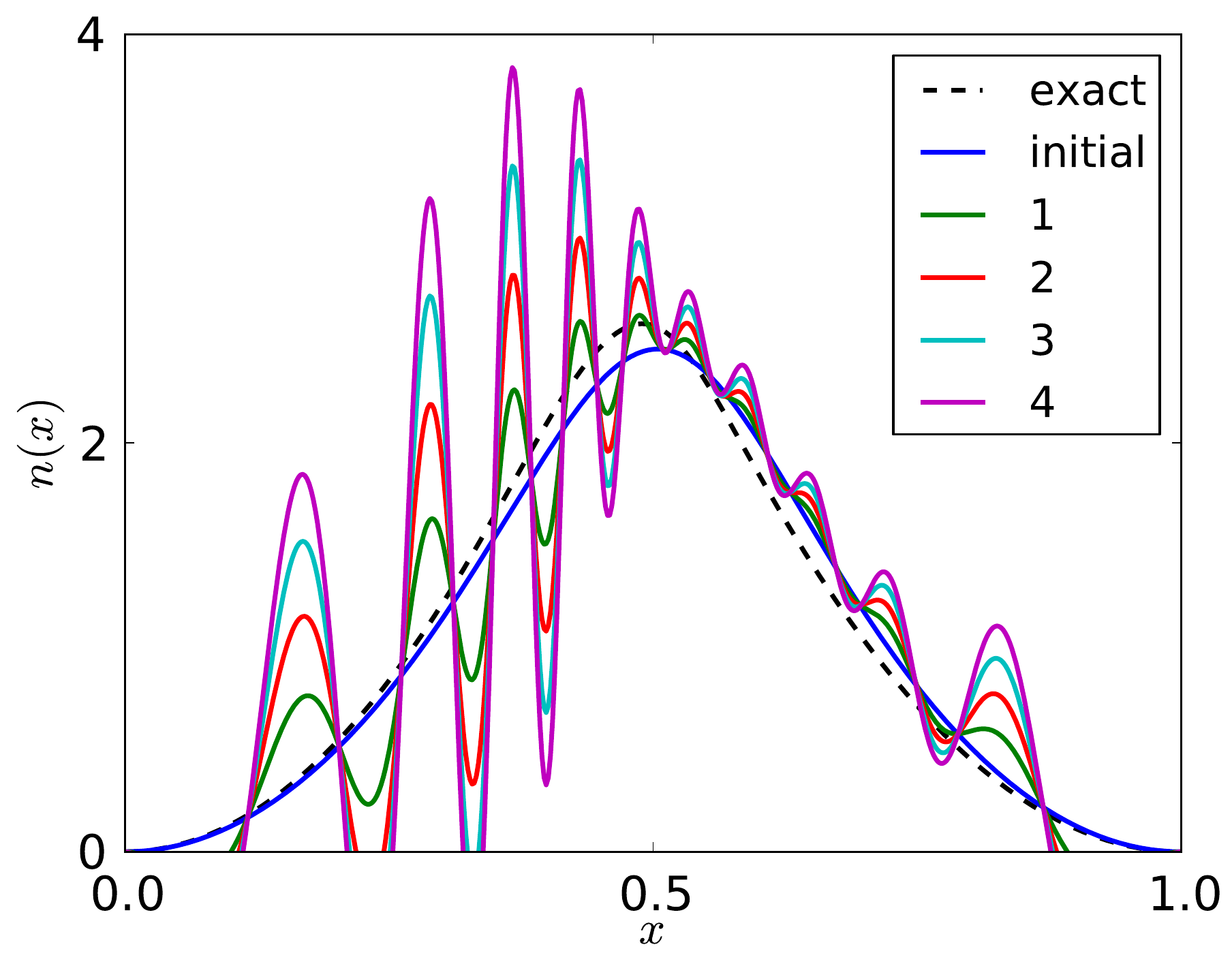}
\caption{The first few steps in a standard gradient descent solving the Euler equation in \Eqref{eq:totEmin}
using our MLA for the KE with $N_T=100$ starting from a sample training density. The dashed line shows the exact self-consistent solution. The noise in the bare functional derivative quickly causes large 
corresponding errors in the density.}
\label{fig:noproj_dens}
\end{figure}

\begin{figure}[tb]
\centering
\includegraphics[width=\figwidth]{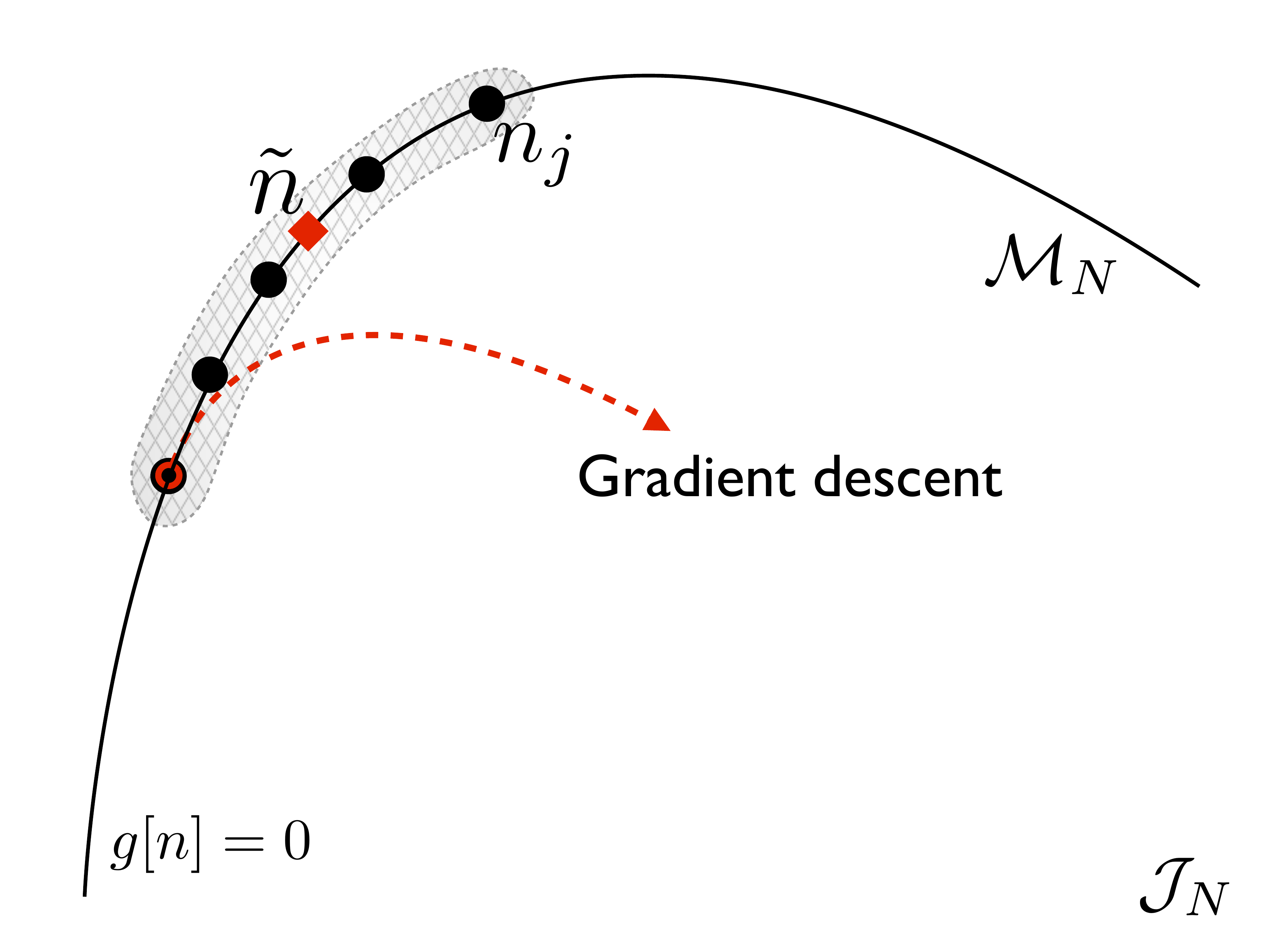}
\caption{Cartoon illustrating the difficulty in solving
for the self-consistent density with our MLA. Pictured are the density manifold $\M_N$ (curved solid line), the training densities $n_j\in\M_N$ (black circles), and the exact self-consistent density $\tilde n$ (red square). Here $g$ is a function that is identically zero on $\M_N$ and positive elsewhere. Thus $\M_N$ is defined implicitly by $g[n]=0$. The shaded area, called the interpolation region, shows where the MLA is accurate. The solution of Eq. \ref{eq:totEmin} via exact gradient descent is given by the red dashed line, which becomes unstable and soon leaves the shaded area.}
\label{fig:densitymanifold}	
\end{figure}

What is the source of this noise?
In general, if the underlying dimensionality of the data
is much less than the dimensionality of $J_N$ (which in this case is essentially infinite), ML will be unable to capture the functional derivative.
The functional derivative contains information on how the KE changes along any direction, but
ML cannot learn this because it only has information in directions in which it has data (i.e. along $\M_N$). 
Fig.~\ref{fig:densitymanifold} illustrates the problem: standard minimization techniques
will rapidly exit the ``interpolation'' region in which the MLA is expected to be accurate.
The MLA is only given information about how the KE changes
along the density manifold $\M_N$. In the many dimensions orthogonal to $\M_N$, the MLA produces an inaccurate derivative (each of these dimensions produces a large relative error since no data exists in these directions; the sum over many dimensions creates a large total error in functional derivative).
A standard gradient
descent will quickly venture off of $\M_N$ into regions of $\J_N$ where the model is guaranteed to fail. \Figref{fig:noproj_dens} shows the deviation of self-consistent density if the search is not constrained to $\M_N$.
To fix this, we further constrain the minimization in \Eqref{eq:totEmin} to stay on $\M_N$. The Euler-Lagrange
minimization for the ground-state density can be expressed as
\ben
\delta \left\{ E[n] - \zeta g[n] \right\} = 0,
\label{eq:totEmin2}
\een
where $g$ is any function that is zero on $\M_N$ and positive elsewhere. Thus 
$g[n] = 0$ implicitly defines the density manifold $\M_N$. Since any $n \in \M_N$ satisfies
the normalization condition, the previous constraint is no longer necessary.
Because the minimizing density (i.e. the ground-state density) is in $\M_N$ and thus satisfies the constraint
$g[n] = 0$, Eq. \ref{eq:totEmin2} gives the same solution as Eq. \ref{eq:totEmin}. Essentially,
we have vastly reduced the domain of the search from $\J_N$ to $\M_N$.
To avoid confusion, we call the minimizing density of this equation the {\em constrained optimal density}.
It may be solved self-consistently in the same sense of solving the standard Euler equation.
However, the $g[n]$ which exactly gives the density manifold is unknown. In the next section,
we develop an approximation which attempts to reconstruct the density manifold from the training
densities.

\begin{figure}[tb]
\centering
\includegraphics[width=\figwidth]{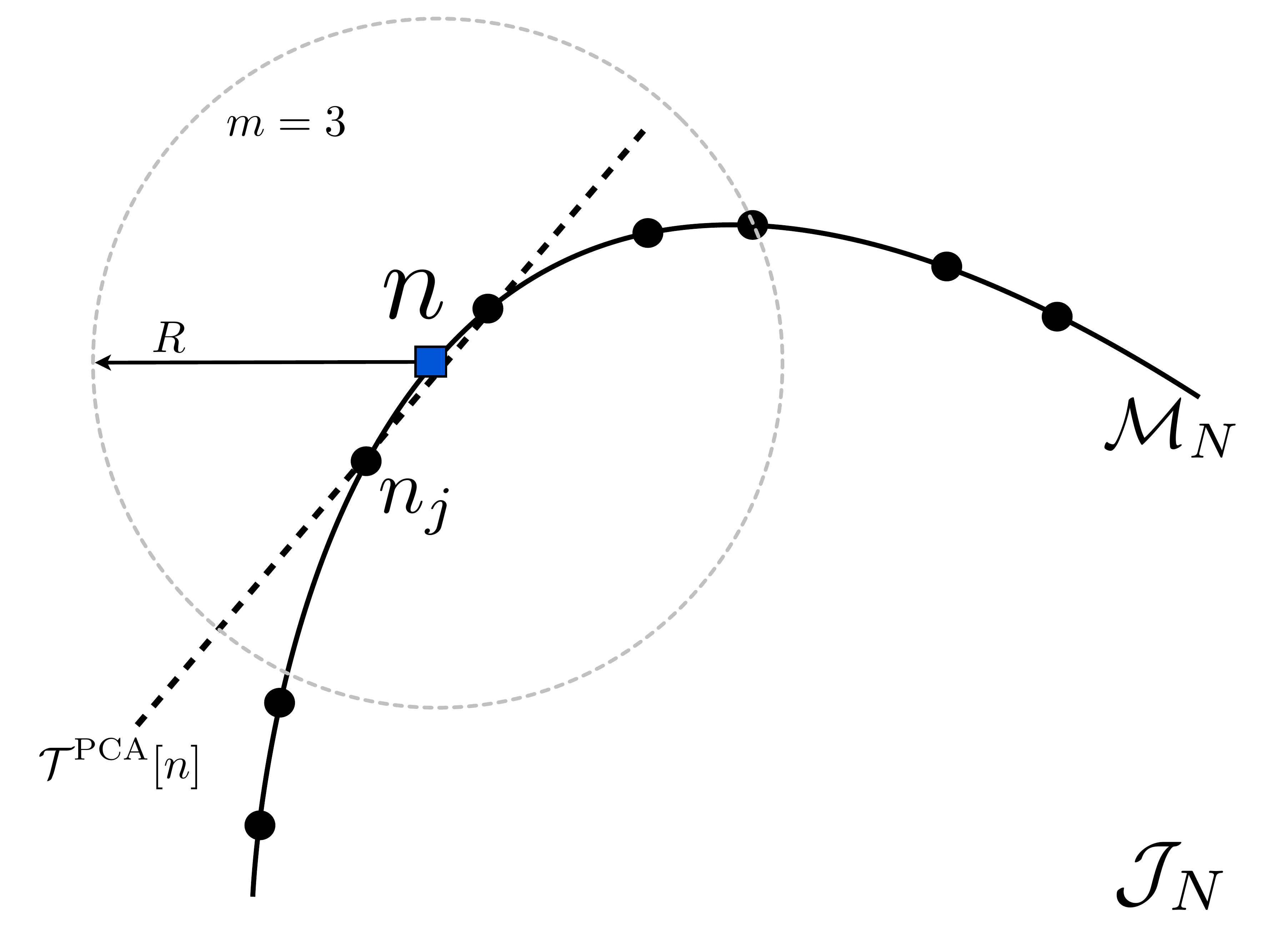}
\caption{Cartoon showing the density manifold $\M_N$ (curved line) that is contained in $\J_N$, the training densities $n_j$ for $j=1,\dots,N_t$ (black circles). Also shown are the density $n \in \M$ (blue square) and the PCA approximation to tangent space of $\M_N$ at $n$, $\T\PCA(n)$ (dashed line). This tangent plane is a local approximation to $\M_N$.}
\label{fig:localPCA}
\end{figure}

\ssec{Manifold reconstruction using principal component analysis}

Our aim is to reconstruct $\M_N$ locally around a given density $n(x)$, which is assumed to be on the density manifold.
A simple approach is
to approximate $\M_N$ as locally linear, using
principal component analysis (PCA) to determine the tangent space empirically from the training densities. This will work as long as
there are enough training densities covering the density manifold.
First, we define a weighted average density around density $n$:
\ben
\bar n(x) = \frac{1}{\Omega} \sum_{j=1}^{N_T} \omega_j n_j(x)
\een
This generalized average is weighted by the function $\omega(\|n - n'\|)$ that only depends on the distance from $n(x)$ to $n'(x)$, $\omega_j = \omega(\|n - n_j\|)$, and $\Omega = \sum_{j=1}^{N_T} \omega_j$. Note that $n'(x)$ refers to the density $n'$ evaluated at $x$ and 
not the derivative of $n$ with respect to $x$.

The locality of the method comes from the choice of $\omega$. For standard PCA, the choice is $\omega(r) = \theta(R-r)$, where $\theta$ is the Heaviside function, and $R$ is the distance from $n'$ to the $m$-th nearest training density. This equally weights the nearest $m$ training densities, and ignores all other training densities. This choice was used in \Ref{SRHM12}.
Here, we choose a slightly smoother weighting function:
\ben
\omega(r)= (1-r/R)\theta(R-r)
\een

Next, PCA is performed by spectral analysis of
the empirical covariance operator~\cite{BO94}, based on the weighted average value around $n(x)$.
We define the centered neighborhood by $\tilde n_j(x) = n_j(x) - \bar n(x)$.
In this problem, densities are represented on a grid with $N_G=500$ points, so let $\bn = (n(x_1), \dots, n(x_{N_G}))^\top$ be the vector representation of $n(x)$.
The covariance matrix $\Gamma \in \mathbb{R}^{N_G \times N_G}$ is
\ben
\Gamma = \frac{1}{\Omega}\sum_{j=1}^{N_T} \omega_j \bn_j \bn_j^\top,
\een
with eigendecomposition
\ben
\Gamma \u_j = \lambda_j \u_j.
\een
The eigenvalues are ordered such that $\lambda_j > \lambda_{j+1}$.
The eigenvectors $\u_j$ are called {\em principal components} (PCs), and give the directions of maximum variance
in the data. We define the variance lost in keeping $d$ PCs as $\eta = 1 - \sum_{j=1}^{d} \lambda_j  \Big/ \sum_{j=1}^{N_G} \lambda_j$.
In this case, there is little to no variance in directions orthogonal to the tangent space of $\M_N$, and maximum variance in directions aligned with the tangent space. Thus, 
the first $d$ PCs form a basis for the tangent space, where $d$ is
the dimensionality of the density manifold (and tangent space). The projection
operator onto this basis is:
\ben
P[n] = \sum_{j=1}^d \u_j \u_j^\top.
\label{eq:PCAprojop}
\een
The tangent space using PCA is given by
\ben
\T\PCA[n] = \{ \bn \,|\, (1 - P[n]) (\bn - \bar\bn) = 0 \}.
\een
Finally, we choose the PCA approximation to the constraint $g[n]$ in \Eqref{eq:totEmin2} as the squared distance from $\bn$ to tangent plane $\T\PCA[\bn]$:
\ben
g\PCA[n] = \| (1 - P[n]) \tilde \bn \|^2.
\een
The PCA approximate density manifold $\M\PCA$ 
is then defined implicitly by $g\PCA[n] = 0$.
The process is illustrated in Fig.~\ref{fig:localPCA}.
In the next section we develop a projected gradient descent
method to solve \Eqref{eq:totEmin2}.

\ssec{Projected gradient descent algorithm}
\label{sect:projGDalgorithm}

\begin{figure}[tb]
\centering
\includegraphics[width=\figwidth]{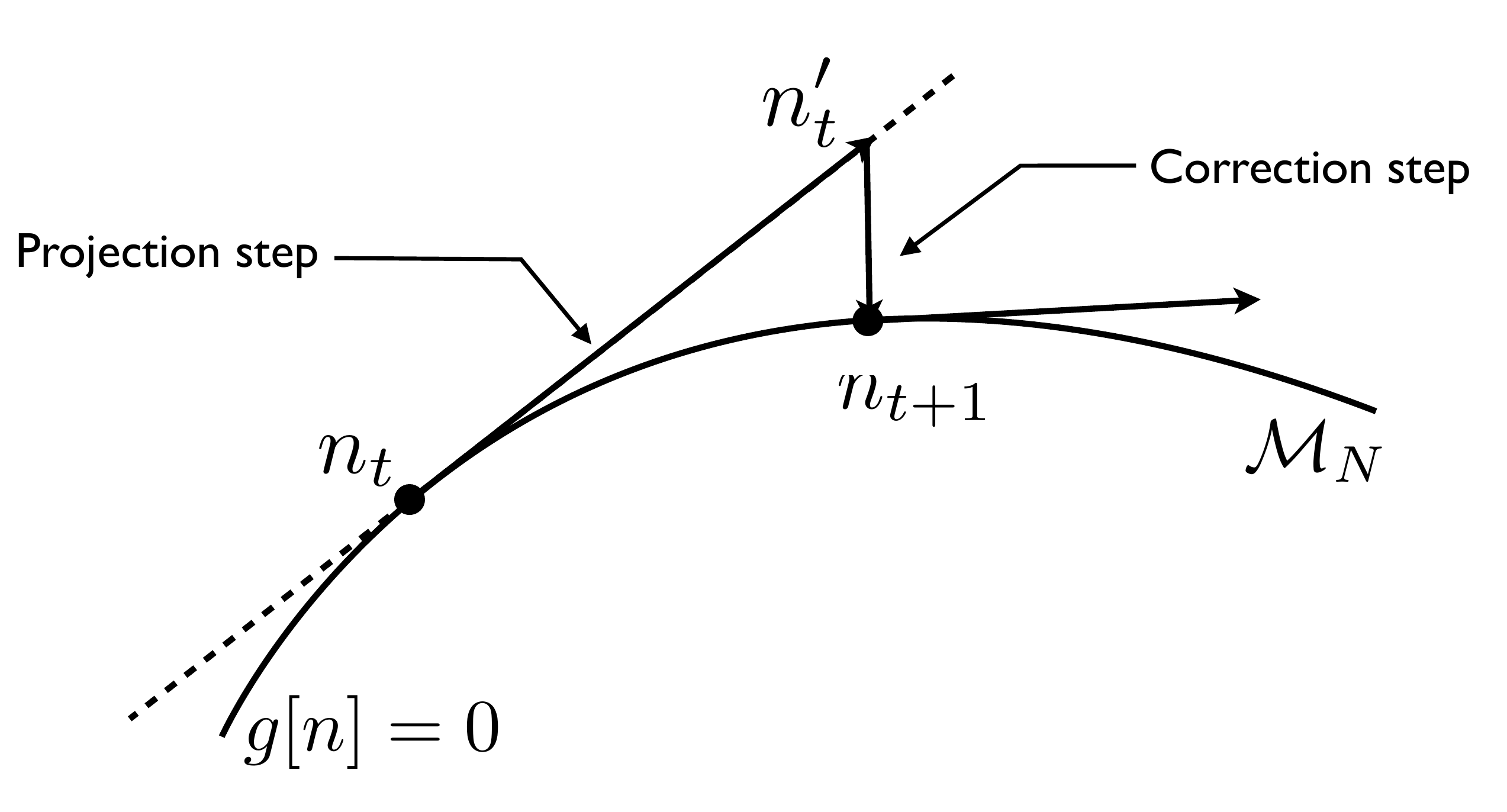}
\caption{Schematic of the projected gradient descent. The functional derivative is projected onto the tangent space of the data manifold $\M_N$ at $n_t$ (dashed line). Next, a step is taken along the projected functional derivative to $n_t'$ in the direction of lower energy. Finally, $g[n]$ is minimized orthogonal to the tangent space to ensure the minimization stays on $\M_N$.}
\label{fig:projectionalgorithm}
\end{figure}

For a given ML approximation to the KE functional,
\ben
E\ML[n] = T\ML[n] + V[n],
\label{eq:MLtotalenergydef}
\een
the algorithm to minimize the functional in \Eqref{eq:totEmin2}
to find a constrained optimal density is as follows (see Fig. \ref{fig:projectionalgorithm}). 
Choose an initial guess for the density, $n_0 \in \M_N$ (e.g., a training density):

\begin{enumerate}
\item Evaluate the functional derivative
\ben
\frac{\delta E\ML[n]}{\delta n(x)} = \frac{\delta T\s\ML[n]}{\delta n(x)} + v(x).
\een
at $n=n_t$.

\item Compute the local PCA projection operator $P[n_t]$ from \Eqref{eq:PCAprojop}.

\item
Project the functional derivative onto the tangent space
 (see Fig. \ref{fig:projectionalgorithm}), and take a step:
\ben
n_{t}'(x) = n_t(x) - \epsilon \hat P[n_t]
\left.\frac{\delta E\ML[n]}{\delta n(x)} \right|_{n=n_t},
\een
where $\epsilon$ is a constant such that $0<\epsilon\leq 1$. If convergence is unstable, reduce $\epsilon$, trading stability for speed of convergence.

\item
To ensure the constraint remains satisfied, we
subtract
the (weighted) mean of the training densities in the local neighborhood:
\ben
n_{t+1}(x) = n'_{t}(x) - (1-\hat P[n'_{t}])(n'_{t}-\bar n[n'_{t}]).
\een

\end{enumerate}

We iterate these steps until
convergence is achieved.
We measure convergence by setting a maximum iteration step and tolerance threshold.
If the total energy difference is smaller than tolerance within max iteration step, the density
 is converged. If no solution is found, $\epsilon$ is reduced.

\ssec{Errors on constrained optimal densities}
\label{error_sc}

With this new constrained minimization procedure via a projected gradient
descent, we solve for the constrained optimal density for 
each test sample. We report the errors in the total energy and KE relative
to the exact density in \Tabref{t:ML:Tperf:short}.
In general, we expect these errors to be worse on the MLA evaluated on
exact densities---by roughly a factor of 10. However, errors on
constrained optimal densities decrease at the same rate with more training data,
so an accuracy of 1 kcal/mol in KE is achieved with 150 training samples for $N=1$,
now on constrained optimal densities.
Additionally, errors are of similar magnitude for multiple particles.
In the last row of \Tabref{t:ML:Tperf:short}, we combine the training data from
each $N$ (100 training densities per $N$ value) into one model. This combined
MLA gives roughly the same error as each individual model. This is because,
due to the locality of the Gaussian kernel, the training densities from each $N$ are well
separated (orthogonal in feature space) and the individual models are unaffected.

\begin{figure}[tb]
\centering
\includegraphics[width=\figwidth]{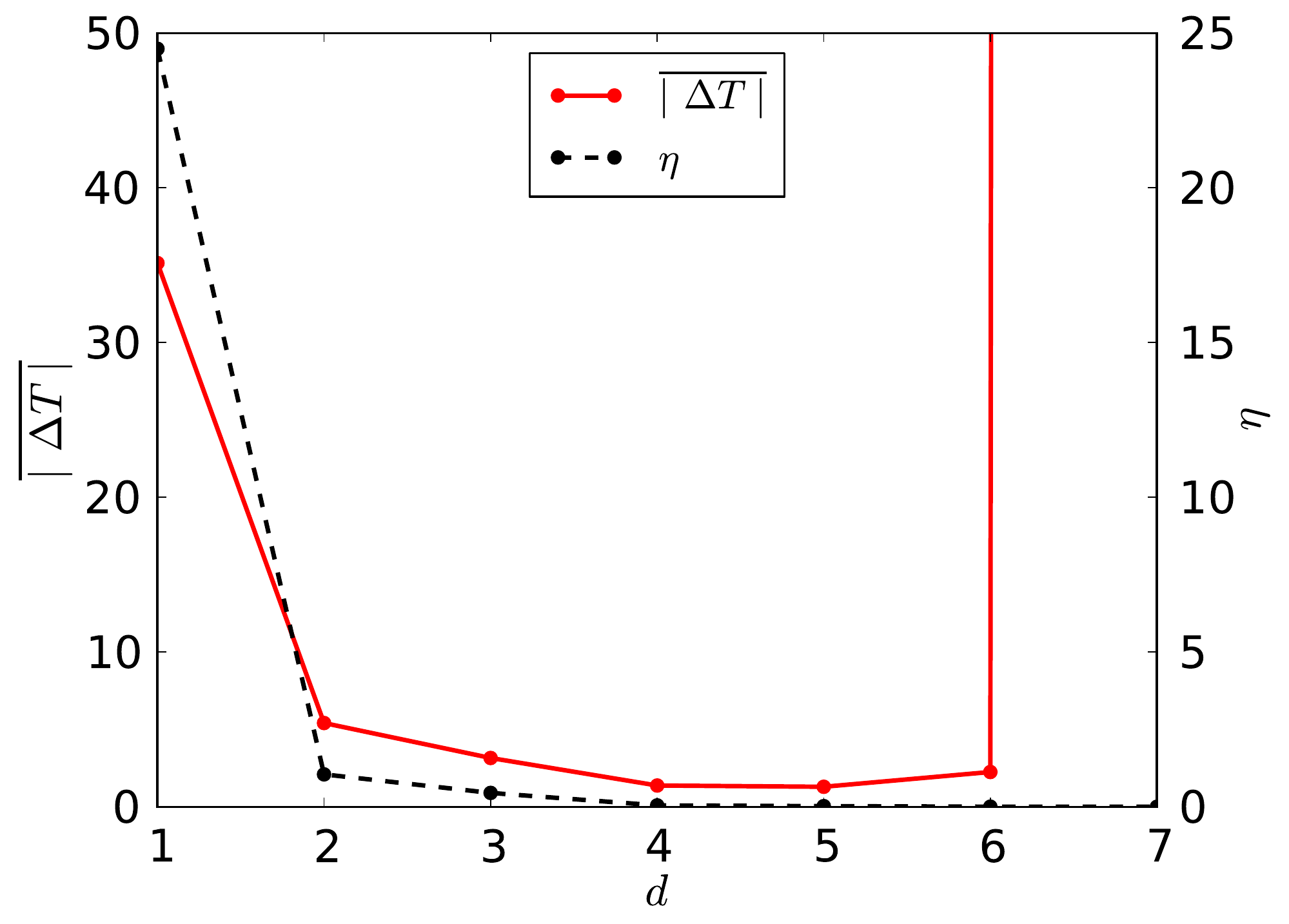}
\caption{The MAE, $\overline{|\Delta T|}=\overline{|T\ML[\tilde n]-T[n]|}$, evaluated on 100 constrained optimal densities (in kcal/mol) compared with the variance lost $\eta$ as a function of 
the number of PCs $d$ in the PCA projection, with $m=20$ nearest neighbors.}
\label{fig:WPCA_para}
\end{figure}

In the projected gradient descent, there are two PCA parameters that 
must be chosen:
$m$, the number of nearest neighbors and $d$, the number of PCs to form the projection
operator. 
\Figref{fig:WPCA_para} shows the MAE evaluated by the constrained optimal density and variance lost as a function of the number of PCs $d$ with $m=20$.
The MAE decreases initially as $d$ increases as more PCs capture the local structure of the
density manifold. As can be seen, $d=4$ or 5 gives an optimal reconstruction of the tangent
space of the manifold. As $d$ increases further, the noise that was removed is re-introduced
into the projection, causing the gradient descent algorithm to fail. For $d=7$, many of the
constrained searches do not converge.
\Tabref{t:MLoptimized_error} reports the errors of the model evaluated on constrained optimal densities for $N_T=40$ and $N_T=100$, giving a rough optimization of the PCA parameters.
Although the potential which generates $\M_N$ has 9 parameters in this case, we
observe that the optimal choice of $d$ is only 4. This is because the data used 
to build the model is only a small fraction of $\M_N$. If we do not sample all relevant directions
on $\M_N$, then the model cannot learn the functional derivative in those 
directions. The PCA projection will compensate by removing those directions. Thus, the
effectiveness of our method depends on the sampling on the manifold.

\begin{table}[tb]
\vspace{0.4cm}
(a)
\begin{tabular}{|c|D{(}{(}{7.5}|D{(}{(}{5.5}|D{(}{(}{5.5}|}
\hline
\backslashbox{$d$}{$m$} & 10 &  20 &  30 \\
\hline
2  & 12\,(98) & 15\,(100) & 24\,(100) \\
3  & 12\,(100) & 16\,(100) & 22\,(100) \\
4  & 12\,(98) & 15\,(100) & 25\,(100) \\
5  & 23000\,(18) & 130\,(27) & (0) \\
\hline
\end {tabular}
\\ \vspace{0.4cm}
(b)
\begin {tabular}{|c|D{(}{(}{4.4}|D{(}{(}{4.4}|D{(}{(}{4.4}|D{(}{(}{4.4}|}
\hline
\backslashbox{$d$}{$m$} & 10 & 20 & 30 & 40\\
\hline
3  & 4.1\,(99) & 3.2\,(100) & 2.7\,(99) & 2.8\,(100) \\
4  & 1.7\,(100) & 1.4\,(100) & 1.4\,(100) & 1.7\,(100) \\
5  & 1.6\,(100) & 1.3\,(100) & 1.5\,(100) & 2.0\,(100) \\
6  & 1.7\,(93) & 2.1\,(100) & 1.7\,(100) & 2.2\,(100) \\
\hline
\end {tabular}%
\caption{The error in the KE in kcal/mol evaluated on constrained optimal densities using 100 densities for testing, with $N=1$ for (a) $N_T=40$ and (b) $N_T=100$. 
The percentage of converged optimal densities is given in parentheses.
Here $m$ is the number of nearest neighbor densities used in PCA
and $d$ is number of PCs used in the projection.}
\label{t:MLoptimized_error}
\end{table}

\sec{Conclusion}

In this work, we have explored in much greater detail the methods presented in \Ref{SRHM12},
in which ML methods were used to directly approximate the KE of a quantum system 
as a functional of the electron density, and used
 this functional in a modified orbital-free DFT to obtain
highly accurate self-consistent densities and energies.

We used a simple model as a proof of principle, to investigate how standard methods
from ML can be applied to DFT. In particular,
we tested a variety of standard kernels used in ML, and have found that the Gaussian kernel
gives the lowest errors (the Cauchy kernel also achieves similar performance). 
All cross validation schemes that were tested gave similar predictions of
hyperparameters that achieved low generalization error on the test set.
Our results highlight the importance of an appropriate choice of kernel, as
some of the kernels tested gave strikingly bad performance. 
With the construction of the $L^2$ norm that was used in the kernels, 
the method is basis set independent (as long as a complete basis is used). 
However, the ML method is capable of learning accurate KEs using a sparse
grid (i.e., an incomplete basis). Using a sparse representation for the density
without losing accuracy would speed up calculations further. 
These results warrant further exploration and will be the subject of future work.

We explained the origin of the noise in the functional derivative and developed
a constrained search over the density manifold via a modified Euler equation, 
effectively projecting out the noise. We also introduced a local approximation to the
manifold using PCA, and solved for constrained optimal densities using a
projected gradient descent algorithm. This worked well for our prototype system,
yielding highly accurate constrained optimal energies and densities.

\acknowledgements

The authors thank for the support from NSF Grant No. CHE-1240252 (JS,
KB). KRM thanks the BK21 Plus Program by NRF Korea, DFG and the Einstein
Foundation. Correspondence to Li Li and K.-R. M\"uller. 

\bibliography{refers}

\clearpage

\end{document}